\documentclass[twocolumn,showpacs,amsmath,amssymb,floatfix,pra]{revtex4}
\usepackage{graphicx}
\usepackage{dcolumn}
\usepackage{bm}

\newcommand{\low}{\operatorname{lowdeg}}
\newcommand{\lcm}{\operatorname{LCM}}
\newcommand{\grcd}{\operatorname{GCD}}
\newtheorem{thm}{Theorem}[section]
\newtheorem{cor}[thm]{Corollary}
\newtheorem{lem}[thm]{Lemma}

\begin{document}

\title{Strong many-particle localization and quantum computing with
perpetually coupled qubits}

\author{L.F. Santos$^{1}$, M.I.~Dykman$^{1}$\footnote{e-mail: dykman@pa.msu.edu}, and M.~Shapiro$^{2}$}
\affiliation{ $^{1}$Department of Physics
and Astronomy, and $^{2}$Department of Mathematics, Michigan State
University, East Lansing, MI 48824}
\author{F.M.~Izrailev}
\affiliation{Instituto de F\'isica,
Universidad Aut\'onoma de Puebla, Puebla 72570, M\'exico }

\date{\today}

\begin{abstract}
We demonstrate the onset of strong on-site localization in a
one-dimensional many-particle system. The localization is obtained by
constructing, in an explicit form, a bounded sequence of on-site
energies that eliminates resonant hopping between both nearest and
remote sites. This sequence leads to quasi-exponential decay of the
single-particle transition amplitude. It also leads to strong
localization of stationary many-particle states in a finite-length
chain. For an {\it infinite} chain, we instead study the time during
which {\it all} many-particle states remain strongly localized.  We
show that, for any number of particles, this time exceeds the
reciprocal frequency of nearest-neighbor hopping by a factor $\sim
10^5$ already for a moderate bandwidth of on-site energies. The
proposed energy sequence is robust with respect to small errors. The
formulation applies to fermions as well as perpetually coupled
qubits. The results show viability of quantum computing with
time-independent qubit coupling.

\end{abstract}

\pacs{03.67.Lx,72.15.Rn,75.10.Pq,73.23.-b}

\maketitle

\section{Introduction}

Disorder-induced localization has been one of the central problems of
condensed matter physics, starting with the Anderson paper
\cite{Anderson}. It has also attracted much attention in other
physical contexts, quantum chaos being an example
\cite{chaos_reviews1,chaos_reviews2}. Recent interest in quantum
computing has further emphasized its importance and allowed looking at
it from a somewhat different perspective.

In many proposed physical implementations of a quantum computer (QC)
the qubit-qubit interaction is not turned off
\cite{liquid_NMR,Makhlin01,mark,mooij99,Yamamoto02,Nakamura03,van_der_Wiel03}. The
interaction may lead to excitation hopping from one qubit to another.
However, control and measurement should be presumably performed on
individual qubits. Therefore it is essential to prevent excitation
transfer between operations. This makes localization a prerequisite
for quantum computing with perpetually coupled qubits. Several
approaches to quantum computing with perpetually coupled qubits have been
proposed recently \cite{Zhou02,Benjamin03,Berman-01}

In a multi-excitation system like a system of interacting electrons or
a QC, inter-site (or inter-qubit) transitions are a many-body effect,
they involve several excitations. In spite of the broad interest in
the problem of many-particle localization, only a limited number of
analytical results has been obtained \cite{Altshuler03}. Numerical
results are also limited: classical computers do not allow studying a
large number of particles, because the Hilbert space is exponentially
large. On the other hand, QC's with perpetually coupled qubits provide
a unique means for investigating localization in a controllable
setting.

In this paper we study strong on-site many-particle localization. It
implies that each particle (or excitation) is nearly completely
confined to one site. This is a stronger condition than just
exponential decay of the wave function, and it is this condition that
must be met in a QC.

A well-known argument suggests that it is hard to strongly
localize a disordered many-particle system where the on-site
energies are random and uniformly distributed within a
finite-width band \cite{Shepelyansky}. Indeed, consider a state
where particles occupy $N$ sites. For short-range hopping, it is
directly coupled to $\propto N$ other $N$-particle states. With
probability $\propto N$ one of them will be in resonance with the
initial state. For large $N$ this leads to state hybridization
over time $\sim J^{-1}$, where $J$ is the inter-site hopping
integral (we set $\hbar = 1$).

In a QC, the quantity $J$ is determined by the qubit-qubit interaction
and usually characterizes the rate of two-qubit operations. At the
same time, the qubit energies are not random and often can be
individually controlled. It is this control that
makes QC's advantageous for studying many-particle localization, as it
becomes possible to construct a ``disordered'' energy sequence site by
site.

Localization can be considered from two points of view. One is based
on the analysis of {\it stationary} states of a many-particle
system. The other is based on studying the system dynamics. Consider a
state with particles occupying a given set of sites, which is called
an on-site state (or a quantum register). As a result of hopping it
can hybridize with another on-site state with nearly the same
energy. We will study the time it takes for resonant hybridization to
happen, which we call the localization lifetime $t_{\rm loc}$.

In a QC all states have a finite coherence time due to coupling to the
environment and external noise. For successful QC operation, 
delocalization should not occur during this time. For most of the
proposed models of a QC, the coherence time is $\alt 10^5
J^{-1}$. Therefore it is sufficient to have the localization lifetime
$\agt 10^5 J^{-1}$. Such lifetime-based formulation of the
many-particle localization problem is relevant to condensed-matter
systems as well, because of finite decay and decoherence times of
quasiparticles for nonzero temperatures.

Here we construct a bounded sequence of on-site energies in a 1D chain
and show that it leads to a long localization lifetime.  We provide
evidence that it also leads to strong localization of many-particle
stationary states in sections of the chain with length up to 12
sites. In a sense, this is an explicit construction of an efficiently
localizing on-site disorder.

In a QC, on-site excitation energies are interlevel distances of the
qubits. They can often be individually controlled, which makes it
possible to construct an arbitrary energy sequence. However, since the
qubit tuning range is limited, so should be the energy bandwidth. A
smaller bandwidth leads also to a higher speed of quantum gate
operations, particularly if they involve changing qubit energies
\cite{DP01}. Of course, condensed-matter systems always have bounded
bandwidth as well.

In condensed-matter physics, localization by ``controlled'' disorder
has been studied in depth and many interesting results have been
obtained in the context of incommensurate periodic potentials, see
Refs.~\onlinecite{Thouless83,Sokoloff85,Albuquerque03} and papers
cited therein. In contrast to this work, we are interested in strong
on-site localization, and not only for single-, but in the first
place, for many-particle states. The ``potential'' that we propose is
not quasi-periodic, and our analytical techniques, including the
time-dependent formulation for many-body systems, are different from
the methods developed for quasi-periodic potentials.

To strongly localize one particle, the difference between excitation
energies on neighboring sites should be much larger than the hopping
integral $J$.  However, even for nearest neighbor coupling, the
energies of remote sites should also differ to prevent transitions
between them via intermediate (nonresonant) sites. The further away
the sites are, the smaller their energy difference can be. This idea
is implemented in our energy sequence. As a consequence, the
single-particle transition amplitude displays nearly exponential decay
with distance. We show that the decay exponent weakly depends on site
and find rigorous bounds on its value.

For many-particle localization one has to suppress not only
single-particle, but also combined resonances, which involve
simultaneous transitions of several interacting excitations. There is
no known way to eliminate all such resonances in an infinite
system. However, with the increasing ``order'' of the transition,
i.e., the number of involved excitations and/or intermediate virtual
states, the effective hopping integral may quickly fall off, leading
to an increase of the transition time.  Then to obtain a desired
lifetime of a localized state it is sufficient to eliminate resonances
up to a certain order. We explicitly show how to do it up to
fifth order, for our sequence.

In a real system it will be possible to tune the energies only with
certain precision. We study how errors in the energies affect
localization and show that our sequence is stable with respect to
small errors. We also demonstrate that, in terms of strong on-site
localization, even for a small chain the constructed energy sequence
is much better than fully random on-site energies with the
same overall bandwidth.

The paper is organized as follows. In Sec.~II we discuss the
Hamiltonian of coupled qubits and introduce a physically motivated
one-parameter sequence of on-site energies. In Sec.~III one-particle
localization of stationary states is studied and quasi-exponential
decay of the transition amplitude is demonstrated. A rigorous proof of
such decay is provided in Appendix A. The relevant scaling properties
are analyzed in Appendix B. In Sec.~IV the inverse participation ratio
is calculated for many-particle states in a section of a 1D
chain. This ratio can be made very close to one in a broad range of
the parameter of the on-site energy sequence, but it also displays
sharp resonant peaks as a function of this parameter. These peaks are
discussed in Appendix C. The lifetime of localized states is
considered in Sec.~V. A minor modification of the energy sequence
allows one to open a gap in the spectrum of combined many-excitation
transitions up to 5th order, which is sufficient for extremely long
localization lifetime. The role of errors in the on-site energies is
studied, and robustness of the results with respect to these errors is
demonstrated in Sec.~VI. In Sec.~VII a highly symmetric
period-doubling sequence of on-site energies is analyzed along with a
sequence of uncorrelated on-site energies. Both are by far inferior,
in terms of localization, to the sequence discussed in
Secs.~II-V. Sec.~VIII contains concluding remarks.

\section{The model}

The problem of localization can be formulated in a similar way for
one-dimensional systems of interacting fermions and spins. The
formulation also applies to qubits during the time when no gate
operations and measurements are performed, i.e., there are no
time-dependent fields that would modulate the qubits.

The relation between spin and qubit systems is simple: qubits are
two-level systems, and therefore can be described by $S=1/2$ spins in
a magnetic field. Then the excitation energy of a qubit becomes the
Zeeman energy of a spin, whereas the qubit-qubit interaction becomes
the exchange spin coupling. Note that the physical interaction itself
may be of a totally different nature, e.g., electric dipolar or
quadrupolar.

For many proposed realizations of QC's the qubit excitation energies
are large compared to the interaction. Then the qubit-qubit
interaction is described by the spin-coupling Hamiltonian of the form
\begin{eqnarray}
\label{spin_int}
H_S= {1\over
2}\sum^{\prime}\nolimits_{n,m}\left[J_{nm}^{xx} (S_n^xS_{m}^x +S_n^y
S_{m}^y)
+J_{nm}^{zz} S_n^zS_{m}^z\right].
\end{eqnarray}
Here, $n,m$ enumerate sites in the 1D spin chain, $z$ is the direction
of the effective magnetic field, and $J_{nm}^{\mu\mu}$ are the
interaction parameters ($\mu=x,y,z$). In mapping the qubit interaction
on the exchange coupling we kept only those terms which, in
the Heisenberg representation, do not oscillate at qubit transition
frequencies. This is why we have set $J_{nm}^{xx}=J_{nm}^{yy}$. In the
case of qubits, the terms $S_n^xS_{m}^x +S_n^y S_{m}^y\equiv
(1/2)(S_n^+S_{m}^- +S_n^- S_{m}^+)$ lead to excitation transfer
between the qubits $n,m$ provided their energies are close. We note
that the spin interaction of the form (\ref{spin_int}) conserves the
number of excitations in the system.

It is convenient to map the spin system onto a system of spinless
fermions via the Jordan-Wigner transformation \cite{Jordan_Wigner}. For
nearest neighbor coupling, the Hamiltonian of the fermion system
becomes
\begin{eqnarray}
\label{hamiltonian_fermions}
&& H=H_0+ H_i,\\
&&H_0=\sum\nolimits_n\varepsilon_na_n^{\dagger}a_n+ {1\over
2}J\sum\nolimits_n\bigl( a_n^{\dagger}a_{n+1}+a_{n+1}^{\dagger}a_n\bigr),
\nonumber\\
&&H_i= J\Delta\sum\nolimits_n
a_n^{\dagger}a_{n+1}^{\dagger}a_{n+1}a_n.\nonumber
\end{eqnarray}
Here, $a^{\dagger}_n, a_n$ are the fermion creation and annihilation
operators; the presence of a fermion on site $n$ corresponds to the $n$th
spin being excited. The parameter $J= J_{n\,n+1}^{xx}$ is the fermion
hopping integral. The parameter $J\Delta = J_{n\,n+1}^{zz}$ gives the
interaction energy of fermions on neighboring sites. If the coupling
of the underlying spins is isotropic, we have $\Delta =1$.  The
on-site fermion energies $\varepsilon_n$ are the Zeeman energies of
the spins (excitation energies of the qubits) counted off from the
characteristic central energy which is the same for all spins inside
the chain. For concreteness we set $J,\Delta> 0$.

Localization, and in particular weak localization, is often described
in terms of the decay of the wave functions of stationary states at
large distances. In contrast, here we are interested in strong on-site
localization. It is determined by the short-range behavior of
stationary states and corresponds to confinement of each particle
to one or maybe a few neighboring sites.

Strong localization can be conveniently characterized by the inverse
participation ratio (IPR), which shows over how many sites the wave
function spreads. For an $N$-particle wave function
$|\psi_{N\lambda}\rangle$ ($\lambda$ enumerates the stationary states) the
IPR is given by the expression
\begin{equation}
\label{IPR}
I_{N\lambda}=\left(\sum\nolimits_{n_1<\ldots<n_N}\bigl\vert\langle 0|
a_{n_N}a_{n_{N-1}}\ldots
a_{n_1}|\psi_{N\lambda}\rangle\bigr\vert^4\right)^{-1},
\end{equation}
where $|0\rangle$ is the vacuum state. In what follows we will
sometimes use the notation $|\Phi(k_1,k_2,\ldots)\rangle=
a^{\dagger}_{k_1}a^{\dagger}_{k_2}\ldots|0\rangle$ for the on-site
wave function (quantum register) in which sites $k_1,k_2,\ldots$ are
occupied and other sites are empty. The quantity (\ref{IPR}) is also
sometimes called the number of participating components, it shows how
many matrix elements
$\langle\Phi(k_1,k_2,\ldots)|\psi_{N\lambda}\rangle$ are substantial,
i.e., of order of their maximal value with respect to $k_1,k_2,\ldots$.

For fully localized states $I_{N\lambda}=1$. Strong on-site
localization means that $I_{N\lambda}$ is close to $1$ for all states
$\lambda$. In this case both the average IPR
\[\langle I_N\rangle =C_{N\lambda}^{-1}\sum\nolimits_{\lambda}I_{N\lambda}\]
%
and $I_{N\max}=\max_{\lambda}I_{N\lambda}$ are close to one. Here,
$C_{N\lambda}$ is the total number of $N$-particle states; for an
$L$-site chain $C_{N\lambda}=L!/N!(L-N)!$. Smallness of $\langle
I_N\rangle -1$ is a weaker condition, it is an indication of strong
localization of most of the states.

In the opposite limit of extended states we have $\langle
I_{N\lambda}\rangle \sim C_{N\lambda} \gg 1$. A simple example is the
case of one particle in an open chain (a chain with free boundaries)
with $\varepsilon_n =$const. The wave functions of the particle are
sinusoidal, and for an $L$-site chain $\langle I_1 \rangle =
2(L+1)/3$. The mean IPR sharply increases with the number of particles
$N$, for $N \leq L/2$.

\subsection{The on-site energy sequence}

Localization requires that the on-site energies $\varepsilon_n$ be
tuned away from each other. The strategy for choosing the sequence
of $\varepsilon_n$ while keeping the overall bandwidth of the
energy spectrum finite can be as follows.  First, we separate the
energies of nearest neighbors by splitting $\varepsilon_n$ into
two subbands, with even and odd $n$, respectively. The distance $h$
between the subbands should significantly exceed the hopping
integral $J$. Then each subband is further split into two subbands
in order to detune next nearest neighbors. The splitting between
these subbands can be less than $h$. This is because
next-nearest-neighbor hopping occurs via virtual transitions to a
nonresonant nearest-neighbor site, and therefore the effective
hopping integral is $\sim J^2/h$. The procedure of band splitting
should be continued, and higher-order splitting can be smaller and
smaller.

We now introduce a simple sequence of $\varepsilon_n$ that implements
the structure described above. Except for the energy scaling factor
$h$, this sequence is characterized by one dimensionless parameter
$\alpha$. As we show, it can already be efficient in terms of strong
localization. For a semi-infinite chain with $n\geq 1$ we set
\begin{equation}
\label{sequence}
\varepsilon_n={1\over 2}h\left[(-1)^n -
\sum\nolimits_{k=2}^{n+1}(-1)^{\lfloor
n/k\rfloor}\alpha^{k-1}\right],
\end{equation}
($\lfloor \cdot\rfloor$ is the integer part).

Sequence (\ref{sequence}) does not have any simple symmetry. For
example, it is not self-similar: the subband widths do not scale with
the distance between the sites that belong to the same subband (cf.
Sec.~VII). Nor is sequence (\ref{sequence}) quasi-periodic. However,
the coefficients at any given power $\alpha^q$ are repeated with
period $2(q+1)$. This important property is essential for obtaining
analytical results, see Appendix A.

\begin{figure}[h]
\includegraphics[width=3.2in]{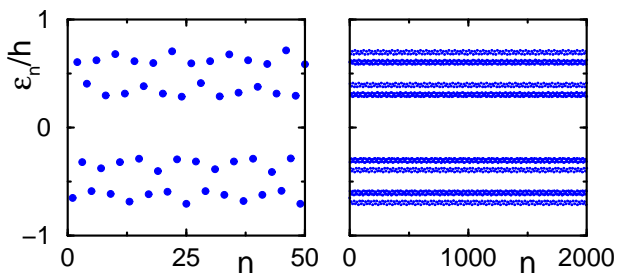}
\caption{(color online) The energies $\varepsilon _n/h$ for
$\alpha=0.3 $. The left panel shows $\varepsilon _n/h$ for the sites
$n=1,2,\ldots,50$. Sites with close on-site energies are spatially
separated.  The right panel shows $\varepsilon _n/h$ for a much longer
array, $n=1,\ldots,2000$. The energy spectrum displays a multisubband
structure, with clearly identifiable 16 subbands in this case.}
\label{fig:energies}
\end{figure}

The energy spectrum (\ref{sequence}) is illustrated in
Fig.~\ref{fig:energies}.  The left panel gives the energies of the
first 50 sites. It is seen that the sites with close $\varepsilon_n$
are spatially separated, whereas the sites with close $n$ are
separated energetically. The multisubband structure of the spectrum is
clearly seen in the right panel. For small $\alpha$, the two major
subbands have width $\approx \alpha h$ and are separated by $\approx
h$. The splitting of higher-order subbands is proportional to higher
powers of $\alpha$. As $\alpha$ increases the subbands start
overlapping, and for $\alpha \agt 0.4$  separation between the
subbands disappears.

As a result of the low symmetry, different subbands in the right panel
of Fig.~\ref{fig:energies} have different numbers of points, i.e.,
$\varepsilon_n$ are not evenly distributed among the subbands. This
turns out to be important for strong many-particle localization. The
case of a symmetric sequence is discussed in Sec.~VII.

An important advantageous feature of sequence (\ref{sequence}) for
quantum computing is that it  is convenient for performing gate
operations. For single-qubit gates, a single particular radiation
frequency can be used to resonantly excite different qubits. It has to
be chosen near the average single-qubit transition frequency (which
corresponds to $\varepsilon = 0$). Then qubits can be selectively
excited by tuning them to this frequency. The transition frequency of
the qubit depends on whether neighboring qubits are excited. This can
be used for implementing a CNOT gate. Alternatively, neighboring
qubits can be tuned in resonance with each other, which will lead to a
two-qubit excitation swap \cite{DP01}.

\section{Single-particle localization: Stationary states}

\subsection{The transition amplitude}

In a 1D system with random on-site energies all single-particle
stationary states are localized, even for weak disorder, and
exponentially decay at large distances. Although sequence
(\ref{sequence}) is not random, the transition amplitudes also display
quasi-exponential decay at large distances provided $J\ll \alpha h$,
as follows from the results of Appendix A. In this paper we are
primarily interested in the short-range behavior. It turns out that a
particle is confined much stronger in the case of sequence
(\ref{sequence}) than in the case of random on-site energies
distributed within the same energy band, see Sec.~VII. The confinement
quickly strengthens with the increasing parameter $\alpha$ once
$\alpha$ exceeds a certain threshold value $\alpha_{\rm th}$.

Spatial decay of single-particle stationary states can be
characterized by the amplitude of a particle transition from site $n$
to site $n+m$. To the lowest order in $J/h$ it has the form
\begin{equation}
\label{amplitude}
K_n(m)= \prod\nolimits_{k=1}^{m}
J/\left|2(\varepsilon_n - \varepsilon_{n+k})\right|.
\end{equation}

For sequence (\ref{sequence}) in the limit of small $\alpha$ the
energy difference $|\varepsilon_{n+m}-\varepsilon_n|$ can be
approximated by its leading term, so it is $\sim h$ for odd $m$ and
$\sim \alpha h$ for odd $m/2$. In general, the larger
is $m$ the higher may be the order in $\alpha$ of the leading term in
$|\varepsilon_{n+m}-\varepsilon_n|$.

The asymptotic behavior of the function $K_n(m)$ for small $\alpha$
and large $m$ can be studied rigorously. The analysis is based on some
results of number theory. It is given in Appendix A.  It shows that
$K_n(m)$ decays with the distance $m$ quasi-exponentially,
\begin{equation}
\label{result} K_n(m)= \alpha^{-\nu|m|}\,(J/2h)^{|m|}.
\end{equation}
The exponent $\nu$ is determined by $\lim \log K_n(m)/m$ for
$m\to\infty$ and depends on $n$. The values of $\nu$ are bound to a
narrow region centered at $\nu= 1$, with $0.89 < \nu < 1.19$. For
estimates one can use $\nu=1$, i.e.,
\[K_n(m)\approx K^{|m|},\qquad K= J/2\alpha h.\]
The decay length of the transition amplitude is then $1/|\ln K|$.

\begin{figure}[h]
\includegraphics[width=2.6in]{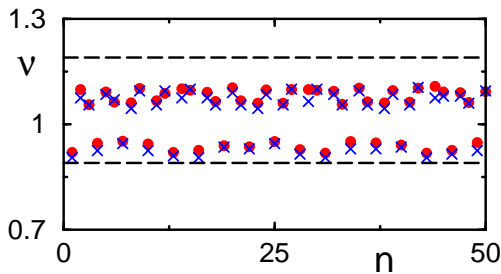}
\caption{(color online). The exponent $\nu$ of the $\alpha$-dependence
of the transition amplitude $K_n(m)$ for the efficient distance $m=200$
(crosses) and $m=1000$ (circles) as a function of site number $n$. The
dashed lines show the analytical limits on $\nu$.}
\label{fig:nu}
\end{figure}

The numerical values of $\nu$ for different $n$ and $m$ are shown in
Fig.~\ref{fig:nu}. They were obtained by keeping the leading term with
respect to $\alpha$ in each energy difference
$\varepsilon_n-\varepsilon_{n+k}$ with $1 \leq k \leq m$. The data are
in excellent agreement with the asymptotic theory.  

We have also studied decay in the ``opposite'' direction, i.e., for
negative $m$ in Eq.~(\ref{amplitude}) (however, $m+n\geq 1$). The
asymptotic expression (\ref{result}) applies in this case, provided
$n,|m| \gg 1$. The numerical data also agree with the theory.

Equation (\ref{result}) gives the tail of the transition amplitude for
$J/2h\alpha\ll 1$. It does not immediately describe strong
single-particle localization, which is determined by the short-range
behavior of the wave function. However, one may expect that strong
localization should occur when $\alpha$ becomes much larger than a
typical threshold value $\alpha_{\rm th} = J/2h$. In fact,
Eq.~(\ref{result}) describes the transition amplitude only when
$\alpha_{\rm th}\ll \alpha \ll 1$. We note that the inequality
$\alpha_{\rm th}< 0.4$, which is necessary to avoid overlapping of the
subbands, is satisfied already when the ratio of the energy bandwidth
to the hopping integral $h/J$ exceeds the comparatively small value
$1.3$.

\subsection{The inverse participation ratio}

\begin{figure}[h]
\includegraphics[width=3.2in]{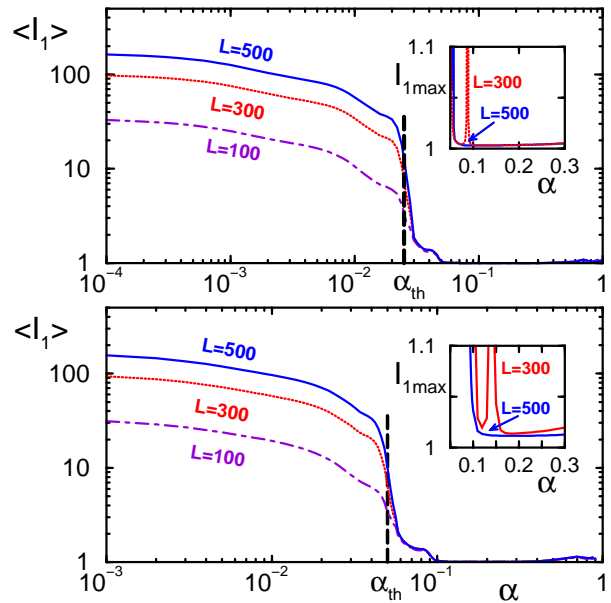}
\caption{(color online). The mean single-particle inverse
participation ratio $\langle I_1\rangle$ vs. $\alpha$ for $h/J=20$
(upper panel) and $h/J=10$ (lower panel). The data refer to three
values of the chain length $L$.  The vertical dashed lines show the
analytical estimate for the threshold of strong localization.  The
insets show the maximal IPR over all eigenstates,
$I_{1\max}\equiv\max_{\lambda}I_{1\lambda}$. It sharply decreases with
the increasing $\alpha$. The peak of $I_{1\max}$ for $L=300$ near
$\alpha = 0.1$ is due to the boundary. Near the minimum over $\alpha$,
we have $I_{1\max}\approx 1.01$ for $h/J=10$, and $I_{1\max}\approx
1.003$ for $h/J=20$. This demonstrates strong single-particle
localization.}
\label{fig:one_excitation}
\end{figure}

A quantitative indication of strong localization of single-particle
stationary states is that $I_{1\lambda}-1 \ll 1$ for all states
$\lambda$. Numerical results on $\langle I_{1}\rangle$ and $I_{1\max}$
as functions of the energy spectrum parameter $\alpha$ for two values
of the scaled bandwidth $h/J$ are shown in
Fig.~\ref{fig:one_excitation}. They were obtained by diagonalizing the
Hamiltonian (\ref{hamiltonian_fermions}) numerically. The data refer to open
chains of three different lengths $L$, with the first site being
always $n=1$ in Eqs.~(\ref{hamiltonian_fermions}),
(\ref{sequence}). The sum over $n$ in the terms $\propto J, J\Delta$
in Eq.~(\ref{hamiltonian_fermions}) ran from $n=1$ to $n=L-1$.

In the fermion Hamiltonian (\ref{hamiltonian_fermions}), the energies
$\varepsilon_n$ differ from the Zeeman energies of spins by $-J\Delta$
inside the chain. On the boundaries this shift is $-J\Delta/2$. To
make the Zeeman energies equal to $\varepsilon_n +$const, with the
constant being the same for all spins, we added the term
$-\frac{1}{2}J\Delta(S_1^z+S_L^z)$ to the spin Hamiltonian
(\ref{spin_int}). Then the numerical results for a finite open chain
equally apply to both spin and fermion systems.

In the limit $\alpha\to 0$, the energies of single-particle stationary states
form two bands centered at $h/2$ for even $n$ and
$-h/2$ for odd $n$.  For $h\gg J$ the widths of these bands are $\sim
J^2/h$. The system is equivalent to two weakly coupled
translationally-symmetric open chains; the band wave functions are
sinusoidal, which gives $\langle I_1\rangle =(L+2)/3$. This agrees
with the value of $\langle I_1\rangle$ for $\alpha\to 0$
in Fig.~\ref{fig:one_excitation}.

For nonzero $\alpha$ the on-site level detuning (\ref{sequence}) breaks
translational symmetry. As $\alpha$ increases, the bands at $\pm h/2$
are split, and more and more subbands are resolved in the energy
spectrum. Respectively, $\langle I_1\rangle$ decreases. It sharply
drops to $\approx 1$ in a narrow region, which can be conditionally
associated with a transition to strong localization. The center of the
transition region gives the threshold value $\alpha_{\rm th}$ of the
parameter $\alpha$. It appears to be independent of the chain length
$L$. The estimate $\alpha_{\rm th}=J/2h$ from Eq.~(\ref{result}) is in
good agreement with the numerical data for different $h/J$.

When $\alpha_{\rm th}\ll \alpha \ll 1$, all stationary states are
strongly localized. Tails of the wave functions are small and limited
mostly to nearest neighbors, which leads to

\begin{equation}
\label{nearest}
I_{1\lambda}-1 \approx J^2/h^2
\end{equation}
to lowest order in $J/h, \alpha$.

For $\alpha \agt 0.4$ the IPR increases above its
minimal value. This happens because the major bands of $\varepsilon_n$
centered at $\pm h/2$ start overlapping.

The minimum of the IPR over $\alpha$ is broad for large $h/J$. Near
the minimum the numerical data in Fig.~\ref{fig:one_excitation} are in
good agreement with the estimate (\ref{nearest}). The agreement
becomes better with increasing $h/J$.

The insets in Fig.~\ref{fig:one_excitation} show that the IPR as a
function of $\alpha$ can have narrow resonant peaks. In the presented
data they occur for the chain of length $L=300$. The peaks are seen
only in $I_{1\max}$, whereas $\langle I_1\rangle$ remains close to 1.
This indicates that only a few on-site states are hybridized with each
other.

The underlying resonance results from a different hopping-induced
shift of the energy levels at the chain edges compared to the bulk.
The analysis of the wave functions shows that the peak corresponds to
a resonance between sites 300 and 296. Because of the hopping, the
energy of site 300 is shifted by $\approx (J/2)^2/h$, whereas for
site 296 this shift is $\approx J^2/2h$. The difference of
$\varepsilon_n$ for these sites is $\sim h\alpha^3$ for $\alpha\ll
1$. Then the peak should occur at $\alpha\approx (J/2h)^{2/3}$, in
good agreement with the data. The effective hopping integral between
the two sites is determined by virtual transitions via intermediate
sites, it is $\sim J^4/16\alpha h^3$. Therefore the width of the peak
with respect to $\alpha$ should be $\propto (J/h)^2$, also in
agreement with the data.

In Appendix B we outline another way of looking at the effect of the
parameter $\alpha$ on localization. Specifically, we study the scaling
relations between $\alpha$ and $J/h$ that follow from the condition
that the IPR takes on a given value.

\section{Many particle localization: stationary states}

\subsection{Many-particle hopping}

The many-particle localization problem is more complicated
than the single-particle one. When the parameter of the inter-particle
interaction $\Delta\neq 0$ (i) the energy levels $\varepsilon_{n}$ are
shifted depending on the occupation of neighboring states,
$\varepsilon_{n} \to \tilde \varepsilon_{n}$, (ii) there occur
combinational many-particle resonances $\tilde\varepsilon_{n_1}+\ldots
+\tilde\varepsilon_{n_k}\approx \tilde\varepsilon_{m_1}+\ldots
+\tilde\varepsilon_{m_k}$, and as a result, (iii) there occur
interaction-induced many-particle transitions that may be resonant
even though single-particle resonances have been suppressed.  Such
transitions may lead to delocalization.

In contrast, the case $\Delta = 0$ corresponds to the $XY$-type
coupling between the underlying spins. In this case the
single-particle results apply to the many-particle system.

To analyze many-particle effects, it is convenient to change from
$a_n^{\dagger},a_n$ to new creation and annihilation operators
$b_n^{\dagger},b_n$ that diagonalize the single-particle part $H_0$ of
the Hamiltonian (\ref{hamiltonian_fermions}),
$a_n=\sum\nolimits_kU_{nk}b_k$. The unitary matrix $\hat U$ is the
solution of the equation
\begin{eqnarray}
\label{U_matrix}
&&\left( U^{\dagger} H_{0} U\right)_{nm}
=\varepsilon^{\prime}_n\delta_{nm},\\
&&(
H_{0})_{nm}=\varepsilon_n\delta_{nm}+{1\over 2}J\left(\delta_{n,m+1}+
\delta_{n+1,m}\right).\nonumber
\end{eqnarray}
Here, $\varepsilon^{\prime}_n$ are the exact single-particle energies,
\begin{equation}
\label{exact_single}
U^{\dagger}H_0U = \sum\nolimits_n \varepsilon^{\prime}_nb_n^{\dagger}b_n.
\end{equation}

For $\alpha \gg \alpha_{\rm th}$ and $J\ll h$, when single-particle
states are strongly localized, the energies $\varepsilon_n^{\prime}$
are close to the on-site energies $\varepsilon_n$. To leading order in
$J/h$, $\alpha$  we have

\begin{equation}
\label{renormalization}
\varepsilon_n^{\prime}-\varepsilon_n \approx
{J^2\over 2h}\left[(-1)^n + {1\over 2} (-1)^{\lfloor n/2
\rfloor}\alpha\right].
\end{equation}
The major term in the right-hand side corresponds simply to
renormalization of the characteristic bandwidth of on-site energies
$h\to h+ J^2/2h$.

In terms of the operators $b_n,b^{\dagger}_n$ the interaction part of
the Hamiltonian is
\begin{equation}
\label{H_int}
U^{\dagger}H_iU=J\Delta\sum
V_{k_1k_2k_3k_4}b_{k_1}^{\dagger}b_{k_2}^{\dagger}b_{k_3}b_{k_4},
\end{equation}
where the sum runs over $k_{1,2,3,4}$, and
\begin{equation}
\label{V4}
V_{k_1k_2k_3k_4} =
\sum\nolimits_pU^*_{pk_1}U^*_{p+1\,k_2}U_{p+1\,k_3}U_{pk_4}.
\end{equation}
The Hamiltonian (\ref{H_int}) describes the interaction of the exact
single-particle excitations.

If the single-particle states are all strongly localized,
the off-diagonal matrix elements $U_{nk}$ are small. They are
determined by the decay of the wave functions, and therefore fall off
exponentially with increasing $|k-n|$. From Eq.~(\ref{result}) we have
$U_{nk}\sim K^{|k-n|}$ for $|k-n| \gg 1$. At the same time, the
diagonal matrix element is $U_{nn} \approx 1$.

Therefore, for strong single-particle localization the major terms in
the matrix $V_{k_1k_2k_3k_4}$ are those with $\varkappa=0$, where
\begin{equation}
\label{varkappa}
\varkappa=\min_p(|k_1-p|+|k_2-p-1|+|k_3-p-1|+|k_4-p|).
\end{equation}
These terms lead to an energy shift $\propto J\Delta$ of the states
depending on the number of particles on neighboring sites.

The meaning of the parameter $\varkappa$ (\ref{varkappa}) can be
understood by noticing that the terms $\propto V_{k_1k_2k_3k_4}$ in
Eq.~(\ref{H_int}) describe two-particle inter-site transitions
$(k_4,k_3)\leftrightarrow (k_2,k_1)$ of the renormalized fermions. For a given
transition, $\varkappa$ is simply the number of virtual steps that
have to be made by the original fermions. The steps are counted off
from the configuration where two such fermions occupy neighboring
sites, and each step is a transition by one of the fermions to a
nearest site. In other words, the original fermions go first from
sites $(k_4,k_3)$ to sites $(p,p+1)$ and then to $(k_2,k_1)$ (we
assume for concreteness that $k_3>k_4$ and $k_1>k_2$); the value of
$p$ is chosen so as to minimize the number of steps.

To make the meaning of $\varkappa$ even more intuitive we give
examples of some $\varkappa=4$ transitions. For the initial and final
states $(n,n+1)$ and $(n-2,n+3)$ one of the sequences of steps of the
original fermions is $(n,n+1)\to (n,n+2)\to (n-1,n+2)\to (n-1,n+3)\to
(n-2,n+3)$, whereas for the initial and final states $(n,n+2)$ and
$(n-1,n+3)$ one of the sequences is $(n,n+2)\to (n,n+1) \to (n-1,n+1)
\to (n-1,n+2) \to (n-1,n+3)$ [the energy denominators must be
obtained directly from Eqs.~(\ref{U_matrix}), (\ref{V4})].

It follows from the above argument that, for large $\varkappa$ and
$\alpha\gg \alpha_{\rm th}$,  the transition matrix element
\begin{equation}
\label{V_vs_K}
V_{k_1k_2k_3k_4} \sim
K^{\varkappa} \qquad {\rm for} \qquad \varkappa \gg 1.
\end{equation}
The transitions of renormalized fermions are not limited to nearest
neighbors. However, from Eq.~(\ref{V_vs_K}), the amplitudes of
transitions over many sites are small and rapidly decrease with the
number of involved virtual steps. The hopping integral for the
renormalized fermions is $VJ\Delta$.

In higher orders of the perturbation theory, the interaction
(\ref{H_int}) leads also to many-particle transitions. The overall
transition amplitude is determined by the total number of involved
virtual single-particle steps.

In order to localize many-particle excitations, one has to suppress
combinational many-particle resonances keeping in mind that, for
localization, the effective hopping integral must be smaller than the
energy detuning of the initial and final on-site states. Because of
the large number of possible resonances, we do not have an analytical
proof of many-particle localization for our energy sequence
(\ref{sequence}). Instead we used numerical analysis, as described in
the next section, which enabled us to demonstrate strong localization
of stationary states in a chain of a limited size.

\subsection{The many-particle inverse participation ratio}

As mentioned earlier, a good indicator of strong localization, which
applies to both single- and many-particle stationary states, is
closeness of the IPR to one. In this section we present numerical data
on the IPR obtained by diagonalizing the Hamiltonian
(\ref{hamiltonian_fermions}) in the presence of several excitations.
The Hamiltonian is a sparse matrix, which is separated into uncoupled
blocks with different numbers of excitations.  We focused on the block
in the middle of the energy spectrum. For a given chain length $L$, it
has $L/2$ excitations and therefore the largest number of states. This
is the worst case, in terms of localization. Chains with $L=10, 12$,
and 14 were studied. The results were similar. We present the data for
$L=12$, in which case the total number of states is 924.

\begin{figure}[h]
\includegraphics[width=3.0in]{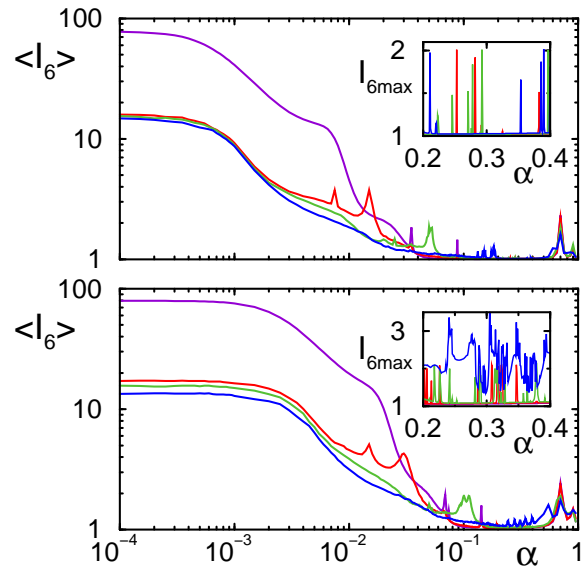}
\caption{(color) The IPR for $6$ excitations on the first 12 sites of
the chain (\protect\ref{sequence}). The reduced bandwidth of the
energy spectrum is $h/J=20$ (top panel) and $h/J=10$ (lower
panel). The purple, red, green, and blue curves refer to the coupling
parameter $\Delta=0,0.3,1$, and 3, respectively. The peaks of $\langle
I_6\rangle$ for $\Delta=0$ are a single-particle boundary effect. The
insets show the maximal $I_6$ for $\Delta\neq 0$. Sharp isolated peaks
of $I_{6\max}$ vs $\alpha$ result from the hybridization of resonating
on-site many-particle states. }
\label{fig:many_loc}
\end{figure}

The IPR as a function of the dimensionless parameter $\alpha$ of the
on-site energy sequence (\ref{sequence}) is shown in
Fig.~\ref{fig:many_loc}. The results refer to two values of the
dimensionless ratio of the hopping integral to the interband distance
$J/h$ and several values of the dimensionless parameter of the
particle interaction $\Delta$.

We start the analysis with the region $\alpha \to 0$, where the
on-site energies (\ref{sequence}) alternate between $\pm h/2$. For
large $h/J$ the single-particle energies form bands of width
$J^2/2h$. In the neglect of mixing of these bands, the many-particle
wave functions can be found using the Bethe ansatz. For $\Delta = 0$
the many-particle energy spectrum consists of bands that are
determined by the number of particles in each of the two
single-particle bands. Because the states are of the plane-wave type,
the IPR is large, with $\langle I_6\rangle \approx 78$ for $h/J=20$
and $L=12$, see Fig.~\ref{fig:many_loc}.

For $\alpha \to 0$ the IPR decreases with the increasing
parameter of the particle interaction $\Delta$. This happens, because
for large $\Delta \gg J/h$ (but $\Delta \ll h/J$) the energy bands at
$\pm h/2$ split into subbands depending on the number of particle
pairs, triples, etc. on neighboring sites: for example, the energy of
a pair on neighboring sites differs from the energy of a dissociated
pair by $J\Delta$. Since the number of states in a subband is smaller
than in the whole band, such splitting reduces the average IPR.

The decrease of the average IPR with increasing $\Delta$ for
$\alpha\to 0$ and $\Delta \ll h/J$ is seen in Fig.~\ref{fig:many_loc}.
We note that the mere separation (by $h$) of the single-particle
energies of neighboring states is not sufficient for strong
localization.  Inside each subband, the number of resonating states in
a long chain is still very large. Localization requires that not only
nearest neighbor single-particle energies, but also energies of remote
sites be tuned away from each other. This should happen for
sufficiently large values of the parameter $\alpha $ in
Eq.~(\ref{sequence}).

This argument is confirmed by the data in Fig.~\ref{fig:many_loc}.
For given $\Delta$, the IPR decreases as a whole with increasing
$\alpha$ in the region where the single-particle bands are well
separated, $\alpha \lesssim 0.4$. This is the expected consequence of
elimination of energy resonances.

In the region $0.2\lesssim \alpha \lesssim 0.4$, except for narrow
peaks, we have $I_{6\max} \approx 1.09$ for $h/J=10 $ and $I_{6\max}
\approx 1.02$ for $h/J=20 $, for $\Delta \leq 1$. The values of
$\langle I_6\rangle$ are even smaller, 1.04 and 1.01,
respectively. This indicates that, in this parameter range, all states
are strongly localized. For $\Delta =3$ and $h/J=20$ we also have
$I_{6\max}\approx 1.01$ away from the peaks; however, for $h/J=10$ the
IPR becomes larger due to the many-particle resonances, which are
discussed below (see also Appendix C).

A distinctive feature of the many-particle IPR as function of $\alpha$
is the onset of multiple resonant peaks, which can be seen in
Fig.~\ref{fig:many_loc}. They indicate that at least some of the
stationary states are no longer confined to a single set of sites,
with the number of sites equal to the number of particles. The peaks
are due to hybridization of resonating on-site states. The
hybridization occurs when the matrix elements of inter-site
transitions in Eqs.~(\ref{H_int}), (\ref{V4}) exceed the energy
difference of the states.

For $0.2 \lesssim\alpha\lesssim 0.4$, i.e., in the region of strong
localization, and for $h/J=20$ and chosen $\Delta \leq 3$ we found
that only two on-site states could become strongly hybridized in the
section of the chain with $1\leq n \leq 12$. Hybridization of a larger
number of states was weaker. A consequence of hybridization of utmost
two states is that $I_{6\max}\lesssim 2$. For $h/J=10$ and $\Delta =3$
the interstate coupling (\ref{V4}) is stronger, and as a result three
states can be strongly hybridized and a few more can be weakly
admixed, leading to $I_{6\max}\sim 3$ at resonant $\alpha$.

Because the interaction is two-particle, the strongest peaks of
$I_{\max}$ come from resonances between on-site energies of two
particles. They occur when the energy difference
\begin{equation}
\label{pair_diff}
\delta\varepsilon
=|\varepsilon_{k_1}+\varepsilon_{k_2}-
\varepsilon_{k_3}-\varepsilon_{k_4}|
\end{equation}
is close to $MJ\Delta$ with $M=0,1,2$. Strictly speaking, we should
use exact single-particle energies $\varepsilon^{\prime}_n$
(\ref{exact_single}) instead of $\varepsilon_n$ in
Eq.~(\ref{pair_diff}), but the difference between these energies is
small, see Eq.~(\ref{renormalization}), and it leads to a small shift
of the positions of the resonances as functions of $\alpha$. An
explanation of the positions and widths of the narrow peaks of
$\langle I_6\rangle$ and $I_{6\max}$ seen in Fig.~\ref{fig:many_loc}
is given in Appendix C.

\subsection{Broad-band two-particle resonances}

A special role in the problem of many-body localization is played by
two-particle resonances that are not selective in $\alpha$, i.e.,
exist in a broad range of $\alpha$. For these broad-band resonances,
the total energy difference between the initial and final on-site
states is small, $\delta \varepsilon \ll J$ even for small $J/h$. They
emerge already in the second order in $J/h$, when only two
single-particle steps to neighboring sites are required, $\varkappa =
2$. The resonating on-site states are pairs $(n,n+1)$ and $(n-1,n+2)$,
i.e.,
\begin{equation}
\label{broad}
\varepsilon_n
+\varepsilon_{n+1} \approx \varepsilon_{n-1}+ \varepsilon_{n+2}.
\end{equation}
If $n$ and $n+2$ are prime numbers, the energy difference $\delta
\varepsilon=|\varepsilon_n +\varepsilon_{n+1}- \varepsilon_{n-1}-
\varepsilon_{n+2}|\sim \alpha^{n-1}h$ is extremely small for large
$n$.

More generally, the resonance (\ref{broad}) occurs for all
$n=6k-1$ with integer $k$. In this case $\delta
\varepsilon/h\propto \alpha^{\xi}$ with $\xi \geq 4$. Such
$\delta \varepsilon$ is ``anomalously small'' for $\varkappa =2$.
The hopping integral $J^3\Delta/h^2$ becomes larger than $\delta
\varepsilon$ even when we are already deep in the single-particle
localization region, $\alpha\gg \alpha_{\rm th}$. 

As a result of renormalization $\varepsilon_n \to
\varepsilon_n^{\prime}$, the energy difference
$\delta\varepsilon=|\varepsilon_n +\varepsilon_{n+1}-
\varepsilon_{n-1}- \varepsilon_{n+2}|$ is changed. For $n=6k-1$ to
leading-order in $J/h$ the renormalization of $\delta \varepsilon$
does not exceed $\sim \alpha^3(J^2/h)$ or $\sim \alpha^4(J^2/h)$ for
odd and even $k$, respectively (it may also be proportional to higher
power of $\alpha$). The renormalized $\delta \varepsilon$ can still be
much smaller than $J^3\Delta/h^2$, and then the broad-band resonance
persists. More many-particle broad-band resonances emerge for larger
$\varkappa$.

For the section of the chain with sites $1\leq n\leq 12$ the IPR is
not much affected by the broad-band resonances, because even where $n$
and $n+2$ are prime numbers (5 and 7), they are not very large and the
energy difference (\ref{broad}) is not exceedingly small.

The situation changes if we consider other sections of the chain
(\ref{sequence}) of the same length, $n_0\leq n\leq n_0+11$ with
different $n_0$. Here we found that the resonances increase $\langle
I_6\rangle$ up to 1.15 between the peaks, for $h/J=20$, $\Delta=1$, and
for all $\alpha$ in the interval $0.2<\alpha<0.4$.

For a finite chain, the resonances can be eliminated order by order in
$\varkappa$ by shifting the energies of the appropriate qubits. Simple
systematic modifications of the energies that work for $\varkappa \leq
5$ are discussed in the next section, see Eqs.~(\ref{correction_6}),
(\ref{correction_3}). Both modifications bring the IPR back to smaller
values. For example, in all sections of the chain that we studied the
made $\langle I_6\rangle$ and $I_{6\max}$ equal to $\approx 1.01$ and
$\approx 1.02$, respectively, which are the values we had for the
section $1\leq n\leq 12$. This indicates that the localization of
stationary states for the modified energy sequences is very strong.

\section{Lifetime of strongly localized states}

The problem of strong localization can be viewed also from a different
perspective. In the context of quantum computing, it suggests a more
appropriate formulation then the one based on the analysis of
stationary states. It is also relevant for condensed-matter systems at
nonzero temperatures.

First we note that excitations in quantum computers and in
condensed-matter systems have a finite coherence time $t_{\rm
coh}$. For QC's, this time has to be compared with the duration of a
single- or two-qubit operation and measurement. The duration of a
two-qubit operation is of order of the time it takes to resonantly
transfer an excitation between the qubits, which is $\sim J^{-1}$. A
single-qubit operation is often faster; however, the measurement can
sometimes be slower. In most proposed realizations of a QC the
coherence time exceeds the gate operation time by a factor $\lesssim
10^5$.

We define the localization lifetime $t_{\rm loc}$ as the time it takes
for excitations to leave occupied sites. Localization of excitations
is only relevant on times $\sim t_{\rm coh}$.  Then to have strong
localization it suffices that $t_{\rm loc}\gtrsim t_{\rm coh}$. It
follows from the estimate for $t_{\rm coh}$ that the latter condition
is met if
\begin{equation}
\label{inequality_bound}
t_{\rm loc}\gtrsim 10^5 J^{-1}.
\end{equation}
The condition (\ref{inequality_bound}) must be satisfied for all
on-site many-excitation states.  It is this condition that imposes a
constraint on the form of the energy sequence $\varepsilon_n$ in an
infinite many-particle system.

The time $t_{\rm loc}$ is determined by hopping between resonant
on-site states.  It occurs through virtual transitions via
nonresonant sites. For a two-particle resonant transition, the minimal number
of the needed virtual steps is given by the parameter $\varkappa$
(\ref{varkappa}). Then from Eqs.~(\ref{H_int}), (\ref{V_vs_K}) the
hopping integral for a resonant transition $(k_4,k_3)\leftrightarrow (k_2,k_1)$ is
\[J\Delta V_{k_1k_2k_3k_4}
\sim J\Delta K^{\varkappa}\]
for $ \varkappa \gg 1$. Here, $K$ is defined by Eq.~(\ref{result}),
$K\approx J/2\alpha h$, and $K\ll 1$ in the region $\alpha/\alpha_{\rm
th}\gg 1$.

The localization lifetime $t_{\rm loc}$ is determined by the
reciprocal maximal hopping integral for resonating many-particle
states.  In the case of an energy sequence of the type
(\ref{sequence}), up to a fairly high number of virtual steps ($\leq
5$), of interest are resonances between two-particle on-site
states. This applies to systems with an {\it arbitrary} number of
particles; only those transitions matter in which up to two particles
change sites. Indeed, transitions where three particles change sites
emerge in the second order in the two-particle Hamiltonian
(\ref{H_int}), (\ref{V4}), and simple counting shows that they involve
already at least 5 virtual steps.

For resonant two-particle transitions $t_{\rm loc}\sim \min [J\Delta
K^{\varkappa}]^{-1}$. Therefore it strongly depends on the minimal
value of $\varkappa$ for all pairs of resonating initial and final
on-site states, $\varkappa=\varkappa_{\min}$.  To have $t_{\rm loc}J$
that exceeds a given value, we must have an appropriate
$\varkappa_{\min}$. This means that we should eliminate resonances
between all states connected by $\varkappa < \varkappa_{\min}$ virtual
transitions.

A two-particle transition with odd $\varkappa=1,3,5,\ldots$ involves a
change of the total number of occupied sites with even $n$ (and by the
same token, with odd $n$, too). Therefore, for the sequence
(\ref{sequence}) with $\alpha\ll 1$, the energy change in such a
transition is $\delta\varepsilon \sim h$. In this paper we consider
the case of weak to moderately strong coupling, when $\Delta \lesssim
1$. Then $\delta\varepsilon$ significantly exceeds the change of the
interaction energy $J\Delta$, $2J\Delta$ for $h\ll J$. As a result,
resonant two-particle transitions may occur only for even $\varkappa$.

We will modify the sequence (\ref{sequence}) to eliminate 
resonances with $\varkappa =2$ and $\varkappa = 4$. In these cases
$\varkappa_{\min}=4$ and $6$, respectively, leading to the
localization time $t_{\rm loc}\sim J^{-1}K^{-4}$ and $>J^{-1}K^{-6}$.

\subsection{Eliminating second order many-particle resonances}

The potentially resonant transitions with $\varkappa =2$ are
\begin{eqnarray}
\label{2nd_order}
&&(n,n+1)\leftrightarrow (n,n+1\pm 2),\;(n-2,n+1)\nonumber\\
&& (n,n+1)\leftrightarrow (n-1,n+2).
\end{eqnarray}
In the transitions listed in the first line of this equation, one of
the particles in the pair moves by two sites in one or the other
direction, whereas for the transition shown on the second line both
particles move by one site.

The number of occupied nearest sites in the transitions
(\ref{2nd_order}) can change by one or remain unchanged. Therefore the
maximal change of the interaction energy is $J\Delta$. Second-order
resonances will be eliminated if the detuning of the on-site energy
differences $\delta \varepsilon$ for the transitions (\ref{2nd_order})
is
\[\delta\varepsilon > J\Delta.\]
This means that we need a zero-energy gap of an appropriate width in
$\delta \varepsilon$.

We note that this is a sufficient, not the necessary condition. In
principle, it would suffice to have narrow gaps at $\delta\varepsilon
=0$ and $J\Delta$. These gaps should just be broader than the
tunneling matrix element and than the energy shifts due to occupation
of next nearest neighbors. For a specific finite-length section of a
chain this may be more practical. However, here we are interested in
an infinite chain, and we want to demonstrate that even for such a
chain all resonances with $\varkappa < 4$ can be eliminated.

To create the zero-energy gap, sequence (\ref{sequence}) has to be
modified. The modification has to eliminate, in the first place, the
``anomalous'' broad-band resonances for transitions
$(n,n+1)\leftrightarrow (n-1,n+2)$ with $n=6k-1$ discussed
before. This will, of course, also eliminate resonances where $n$ and
$n+2$ are prime numbers. A simple and sufficient modification is a
constant shift of $\varepsilon_n$ for each 6th site,
\begin{equation}
\label{correction_6}
\varepsilon_n^{\rm md} = \varepsilon_n +(h/2)\alpha^{\prime}
\qquad {\rm for}\qquad n=6k,
\end{equation}
while $\varepsilon_n^{\rm md} = \varepsilon_n$ for $n\neq
6k$.

For the modified sequence (\ref{correction_6}), the gap in the on-site
energies for the 2nd-order transitions (\ref{2nd_order}) is
$\delta\varepsilon\sim \alpha^2h, \alpha^{\prime}h/2$ to leading order
in $\alpha$. A more accurate estimate is
$\min\delta\varepsilon/h\approx \alpha^2-\alpha^3,
\alpha^{\prime}/2$. We assume that $\alpha^2\lesssim \alpha^{\prime}\ll
\alpha$, in which case no new resonances are created for the
transitions (\ref{2nd_order}) as a result of the modification
(\ref{correction_6}).

It follows from the above estimate that, for an infinite chain and an
arbitrary number of particles, all resonant transitions with
$\varkappa < 4$ will be eliminated provided $J\Delta/h<
\alpha^2-\alpha^3,\alpha^{\prime}/2$, . Then the localization time
$t_{\rm loc} \sim 10^5J^{-1}$ already for $h/J=30$, $\alpha=0.3$,
$\alpha^{\prime}\approx 0.1-0.2$, and $\Delta \alt 1$.

\begin{figure}[h]
\includegraphics[width=3.0in]{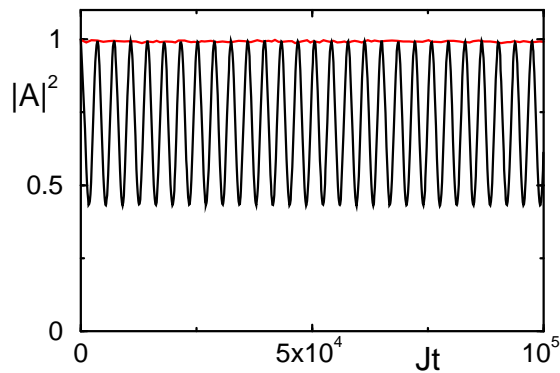}
\caption{(color online) Time evolution of the squared amplitude
$|A|^2$ of the on-site state $|\Phi(416,419,420,422,423,424)\rangle$ in a
12-site section of the chain between $n=415$ and $n=426$. The
oscillating line refers to the original sequence
(\protect\ref{sequence}) with $\alpha=0.25$. The nearly constant line
refers to the modified sequence (\protect\ref{correction_6}) with
$\alpha=0.25, \alpha^{\prime}=0.22$. In both cases $h/J=20$ and $\Delta =1$.}
\label{fig:evolution}
\end{figure}

The effect of the on-site energy modification (\ref{correction_6}) on
localization of a many-particle state is seen from
Fig.~\ref{fig:evolution}. This figure shows the dynamics of the system
prepared initially in the state
$|\Phi(416,419,420,422,423,424)\rangle$. In the case of the original
sequence (\ref{sequence}), this state strongly hybridizes with the
state $|\Phi(416,418,421,422,423,424)\rangle$ for all $\alpha$ of
physical interest, $\alpha < 0.4$. This happens because the difference
of on-site energies $\varepsilon_{419}+ \varepsilon_{420}$ and
$\varepsilon_{418} + \varepsilon_{421}$ is $\delta\varepsilon\sim
\alpha^{418}h$. The hybridization results in oscillations of the
amplitude of the state, as seen from Fig.~\ref{fig:evolution}. For the
modified sequence (\ref{correction_6}) the resonance is eliminated, and
the amplitude remains constant over a time $>10^6J^{-1}$. This
illustrates the onset of strong localization. We note that the
localization time of this particular state turns out to be longer for
the modified sequence (\ref{correction_6}) than the worst-case
estimate given above.

\subsection{Eliminating 4th order resonances}

The localization time is further dramatically increased if $\varkappa=4$
resonances are eliminated. The potentially resonant 4th order  transitions are
\begin{eqnarray}
\label{one_Delta}
&&(n,n+1)\leftrightarrow (n-2,n+3),\nonumber\\
&& (n,n+1)\leftrightarrow (n+2,n+3),\nonumber\\
&&(n,n+3)\leftrightarrow (n-1,n+2),\nonumber\\
&& (n,n+3)\leftrightarrow (n-2,n+1),
\end{eqnarray}
and
\begin{eqnarray}
\label{two_Delta}
&&(n,n+1)\leftrightarrow (n-1,n+4),\nonumber\\
&& (n,n+1)\leftrightarrow (n-3,n+2),\nonumber\\
&&(n,n+1)\leftrightarrow (n,n+1\pm 4),\; (n\pm 4,n+1).
\end{eqnarray}
In the last line of Eq.~(\ref{two_Delta}) we list transitions where
one of the particles in the pair moves by 4 sites, whereas in all
other transitions both particles move away from their sites.

For the modified sequence (\ref{correction_6}), the minimal energy
change in the transitions (\ref{one_Delta}), (\ref{two_Delta}) is
$\min\delta\varepsilon \sim\alpha^3h$, to leading order in $\alpha$,
provided $\alpha^{\prime} \gg \alpha^3$. The value of
$\alpha^{\prime}$ has to be in such a range that the modification
(\ref{correction_6}) does not lead to extra resonances between the
on-site energies for the states (\ref{one_Delta}) and
(\ref{two_Delta}). The ``dangerous'' combinations in
$\delta\varepsilon/h$ are $ |\alpha-\alpha^{\prime}/2|,
|2\alpha-\alpha^{\prime}/2|, |\alpha^2
-\alpha^{\prime}/2|,|2\alpha^2-\alpha^{\prime}/2|$, to leading order
in $\alpha$. We will choose $\alpha, \alpha^{\prime}$ so that all of
them exceed $\min\delta\varepsilon/h\approx \alpha^3$.

\begin{figure}[h]
\includegraphics[width=3.0in]{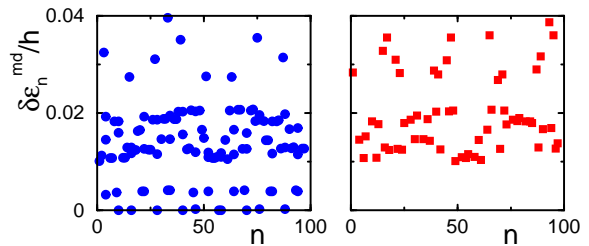}
\caption{(color online) The low-energy part of the two-particle energy
differences $\delta \varepsilon _n /h$ (\protect\ref{pair_diff}) for
all transitions with $\varkappa\leq 5$ in which one of the involved
particles is on the $n$th site ($n> 2$). The data refer to $\alpha
=0.25$. The left panel corresponds to the sequence (\ref{sequence}).
The right panel refers to the modified sequence
(\protect\ref{correction_6}) with $\alpha^{\prime} = 0.22$ and shows
the zero-energy gap in $\delta \varepsilon$.}
\label{fig:correction}
\end{figure}

Fig.~\ref{fig:correction} shows how the modification
(\ref{correction_6}) leads to a zero-energy gap in
$\delta\varepsilon$.  We plot $\delta \varepsilon_n^{\rm md}$ for all
transitions (\ref{one_Delta}), (\ref{two_Delta}) with one of the
particles on site $n$ (with $n > 2$). Therefore we show all
potentially resonant transitions with $\varkappa \leq 5$.

The left panel in Fig.~\ref{fig:correction} shows that, for the
initial sequence (\ref{sequence}), there is practically no gap in the
values of $\delta\varepsilon$ at low energies. The right panel
demonstrates that the correction (\ref{correction_6}) leads to a
zero-frequency gap. The gap depends on the values of $\alpha$ and
$\alpha^{\prime}$.  For the specific parameter values in
Fig.~\ref{fig:correction} we have $\delta\varepsilon/h \geq 0.01$. 

We have checked numerically that the gap persists for a much longer
chain than shown in Fig.~\ref{fig:correction}, with $n$ from 2 to
10,000. This is more than enough to prove that the results apply to an
infinite chain. Indeed, in $\varepsilon_n^{\rm md}$ the terms
$\propto\alpha^q$ with different $q$ are repeated periodically with
period $2(q+1)$. Therefore a sequence of terms with $\alpha^q$ for
$q\leq 6$ is repeated periodically with period equal to twice the
least common multiple of all $q+1\leq 7$, which is $2\times(3\times
4\times 5\times 7)= 840$ (cf. Appendix A). For $\alpha=0.25$ we have
$\alpha^6\approx 2\times 10^{-4}$. The contribution to
$\varepsilon_n^{\rm md}/h$ of all terms of higher degree in $\alpha$
is then $\lesssim 2\times 10^{-4}$. This means that, to accuracy
better than $ 2\times 10^{-4}$, the results on the gap for an infinite
chain will coincide with the results for a chain of 840 sites.

It follows from the discussion above that, for $2J\Delta \lesssim 0.01
h$, all particles will remain localized on their sites for the time
$t_{\rm loc} \sim (J\Delta)^{-1}K^{-6}/\alpha$ [we have taken into
account here that the hopping integral for transitions with $\varkappa
=6$ is limited by $\sim J\Delta (J/2h)^6\alpha^{-5}$ rather than
$J\Delta (J/2h)^6\alpha^{-6}$, as would be expected from the
asymptotic expression (\ref{V_vs_K})]. For $h/J=50$ and $\alpha =0.25$
this gives an extremely long localization time, $t_{\rm loc}J\gtrsim
10^{10}$. However, this estimate requires that the coupling be weak,
$\Delta \lesssim 0.25$ for the used parameter values.

\subsubsection{Extension to stronger coupling}

The previous result can be easily extended to larger $\Delta$ without
increasing $h$. To do this we note that the change of the interaction
energy in transitions (\ref{one_Delta}) is actually limited to
$J\Delta$ rather than $2J\Delta$, which is the case for the transitions
(\ref{two_Delta}). Therefore 
the gap for the transitions (\ref{two_Delta}) should be twice as large
as for the transitions (\ref{one_Delta}).  Two first transitions
(\ref{two_Delta}) have a gap $\gtrsim \alpha^2h$ to leading order in
$\alpha$, but for the last one $\min\delta\varepsilon
=\alpha^3h$. This latter gap may be increased by choosing a somewhat
different modification of the on-site energy spectrum. In contrast to
Eq.~(\ref{correction_6}), we will shift the energy of each 3rd site,
\begin{equation}
\label{correction_3}
\tilde\varepsilon_n^{\rm md} = \varepsilon_n -(h/4)\beta[1+3(-1)^k]
\qquad {\rm for}\qquad n=3k,
\end{equation}
while $\tilde\varepsilon_n^{\rm md}=\varepsilon_n$ for $n\neq 3k$.

Eq.~(\ref{correction_3}) corresponds to shifting $\varepsilon_{3k}$ up
by $\beta h/2$ or down by $\beta h$ depending on whether $k$ is odd or
even, respectively. The parameter $\beta $ should be much larger than
$\alpha^3$, to open a gap $\sim\min(\alpha^2 h, \beta h/2)$ in
$\delta\tilde\varepsilon_n^{\rm md}$ for the transitions
(\ref{two_Delta}). At the same time, it should be chosen so as to
avoid creating new resonances, similar to the modification
(\ref{correction_6}).

It is straightforward to show that the modification
(\ref{correction_3}) leads to a zero-energy gap in
$\delta\varepsilon$. For $\alpha =0.25,\beta = 0.1725$ the gap exceeds
$0.01 h$ for the transitions (\ref{2nd_order}), (\ref{one_Delta}) with
energy change $\lesssim J\Delta$, whereas for the transitions
(\ref{two_Delta}) with energy change up to $2J\Delta$ it exceeds $0.02
h$. This indicates that the results on the localization time $t_{\rm
loc}J\gtrsim 10^{10}$ for $h/J=50$ now apply for $\Delta \lesssim
0.5$. 

As in the previous section, the extremely large localization time
characterizes an infinite chain and an arbitrary number of interacting
particles. We note that both modifications of the original energy
sequence, Eqs.~(\ref{correction_6}) and (\ref{correction_3}), are
obtained analytically, by finding the leading-order terms in the
energy differences for the transitions (\ref{2nd_order}),
(\ref{one_Delta}), and (\ref{two_Delta}). The specific values of the
parameters $\alpha$, $\alpha^{\prime}$, and $\beta$ are used just to
illustrate the order of magnitude of the localization time.

\section{Stability with respect to errors in on-site energies}

In a real system, it will be impossible to implement sequence of
on-site energies (\ref{sequence}) precisely. This is because these
energies contain high powers of the small parameter $\alpha$, while
the precision to which they can be set and/or measured is
limited. Therefore it is necessary to study localization in the
presence of errors in $\varepsilon_n$ and to find how large these
errors can be before they cause delocalization.

We will address this problem by looking at the gap in the energy
differences $\delta\varepsilon$ in the presence of errors in
$\varepsilon_n^{\rm md}$ (\ref{correction_6}). As long as this gap
remains larger than $2J\Delta$ for all resonant transitions with
$\varkappa \leq 5$, the localization lifetime $t_{\rm loc}$ will
remain large. The analysis can be immediately extended to the modified
sequence (\ref{correction_3})  as well.

The effect of errors on the gap can be modelled by adding a
random term to on-site energies, i.e., replacing
$\varepsilon_n^{\rm md}$ with
\begin{equation}
\label{error}
\varepsilon_n^{\rm err} = \varepsilon_n^{\rm md} + {1\over 2}Dhr_n.
\end{equation}
Here, $r_n$ are random numbers uniformly distributed in the interval
$(-1,1)$, and $D$ characterizes the error amplitude. It should be
compared with $\alpha^s$ with different exponents $s \geq 1$. When
$D\sim \alpha^s$ it means that the energies $\varepsilon_n$ are well
controlled up to terms $\sim \alpha^{s-1}$, to leading order in
$\alpha$.

From the above arguments it follows that, for $\alpha \gg
\alpha^{\prime} \agt \alpha^2$ the gap should remain unchanged if
$D\ll \alpha^4$. This is because, for the modified energies
$\varepsilon_n^{\rm md}$, the terms $\sim \alpha^4$ drop out from the
energy differences that we discuss. For $D\sim \alpha^4$ the gap should be
somewhat reduced.  For $D\sim \alpha^3$ it should become significantly
smaller than for $D=0$, and it should ultimately disappear with
increasing $D$.

\begin{figure}[h]
\includegraphics[width=3.0in]{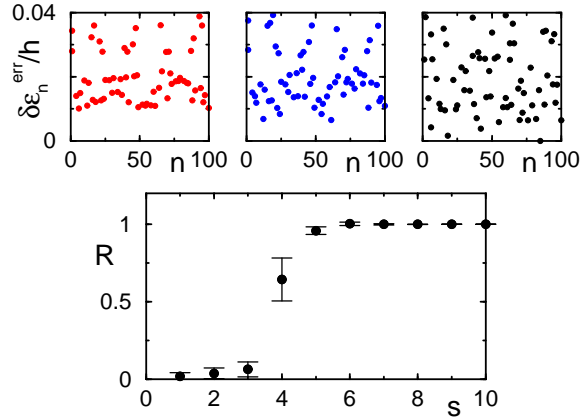}
\caption{(color online) Upper panels: all energy differences $\delta
\varepsilon _n^{\rm err} /h= |\varepsilon_{n}^{\rm
err}+\varepsilon_{n_1}^{\rm err}- \varepsilon_{n_2}^{\rm
err}-\varepsilon_{n_3}^{\rm err}|/h $ for the transitions
(\ref{2nd_order}), (\ref{one_Delta}), (\ref{two_Delta}) that
correspond to the number of intermediate steps $\varkappa \leq 5$. The
data refer to $\alpha =0.25$, $\alpha^{\prime} = 0.22$, and to a
specific realization of the random numbers $r_n$ in
Eq.~(\protect\ref{error}). The boxes from left to right correspond to
the values of the noise intensity $D=\alpha^s$ in
Eq.~(\protect\ref{error}) with $s=5,4$, and 3. Lower panel: the scaled
minimal gap $R$ (\protect\ref{R_ratio}) as a function of the logarithm
$s=\ln D/\ln\alpha$ averaged over 10 realizations of noise. Error bars
show the standard deviation of $R$.}
\label{fig:error}
\end{figure}

Numerical results on the gap $\delta\varepsilon$ as a function of $\log D$
are shown in Fig.~\ref{fig:error}. The gap is calculated for
two-particle transitions with $\varkappa\leq 5$, as in
Fig.~\ref{fig:correction}. In the lower panel the gap is scaled by its
value in the absence of errors,
\begin{equation}
\label{R_ratio}
R=\min_n\delta\varepsilon^{\rm
err}_n/\min_n\delta\varepsilon^{\rm md}_n.
\end{equation}
The data refer to the same $\alpha,\alpha^{\prime}$ as in
Fig.~\ref{fig:correction}. They are in full agreement with the above
estimate.

The results of Fig.~\ref{fig:error} demonstrate that the localization
persists even for relatively large errors in the on-site energies. At
least for the chosen $\alpha$ and $\alpha^{\prime}$, errors in
$\varepsilon_n$ up to $\sim 0.4\%$ (when $D=\alpha^4$) lead to a
change in the width of the energy gap by $\sim 50\%$.

The observed dependence on the noise strength suggests that, in the
presence of noise, sequence (\ref{sequence}), (\ref{correction_6})
can be cut so that the terms $\propto \alpha^s$ with $s> s_{\rm
cutoff}$ are disregarded. The value of $s_{\rm cutoff}$ depends on the
noise, $s_{\rm cutoff} = \ln D/\ln \alpha$. As a result of the cutoff,
the energies $\varepsilon_n$ become polynomials in $\alpha$ of power
$\leq s_{\rm cutoff}$. From Eq.~(\ref{sequence}), these polynomials
are periodic in $n$, with the period determined by twice the least
common multiple of $(2,3,\ldots,s_{\rm cutoff}+1)$. For example, for
$s_{\rm cutoff} = 5$ the period in $n$ is 120 (see also Appendix
A). For such a long period and short-range hopping, excitations will
stay on their sites for a long time compared to $J^{-1}$.

\section{Alternative energy sequences}

Neither the original sequence of on-site energies (\ref{sequence}) nor
its modified versions (\ref{correction_6}), (\ref{correction_3}) were
optimized to maximize the IPR or the localization lifetime. Therefore
it is important to compare them with other sufficiently simple
sequences. This will be done in this section for two natural choices
of $\varepsilon_n$.

\subsection{Period doubling cascade}

A simple way to move sites with close energies far away from each
other is to make the energies form a ``period doubling cascade''
(PDC). It can be described by a one-parameter energy sequence; in what
follows $h$ is the energy scale and $\alpha$ is the parameter, as in
Eq.~(\ref{sequence}).

In the PDC, the on-site energies are first split into two subbands,
with nearest neighbors being in different subbands, but next nearest
neighbors being in the same subband. The subbands differ in energy by
$\alpha^0h$, to leading order in $\alpha$ [this is also the case in
Eq.~(\ref{sequence})]. Each of the subbands is then further split into
two subbands of 4th neighbors. The leading term in the energy
difference of these subbands is $\alpha^1 h$. Each subband is then
split again into two subbands of 8th neighbors. The leading term in
their energy difference is $\alpha^2 h$. This period-doubling process
is then continued indefinitely, for an infinite chain.

The expression that describes the on-site energy sequence for the PDC,
$\varepsilon_n^{\rm PDC}$, can be conveniently written in terms of the
coefficients $j_k(n)=0,1$ of the expansions of site numbers $n$ in base
two,
\[n=\sum_{k=0}^{M(n)-1} j_{k}(n)2^k.\]
Here, $M(n)=1+\lfloor
\log_2 n\rfloor$ is the number of integer digits of $n$ in base 2.

\begin{figure}[h]
\includegraphics[width=3.2in]{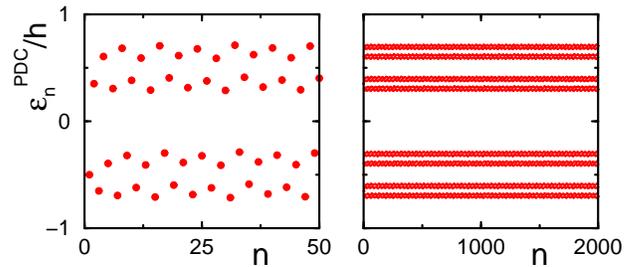}
\caption{(color online) The on-site energies $\varepsilon _n^{\rm
PDC}/h$ (\protect\ref{PDC}) for $\alpha=0.3 $. The left panel shows
the energies for the sites $n=1,2,\ldots,50$. States with close
on-site energies are spatially separated.  The right panel shows
$\varepsilon _n^{\rm PDC}/h$ for a much longer array,
$n=1,\ldots,2000$. The energy spectrum displays a multisubband
structure. Because of the symmetry of sequence
(\protect\ref{PDC}), the number of points in each subband is
approximately the same. }
\label{fig:PDCenergies}
\end{figure}

We set
\begin{equation}
\label{PDC}
\varepsilon_n^{\rm PDC}={1\over 2}h\left[(-1)^n
+\sum_{k=1}^{M(n)-1}(-1)^{j_{k}(n) }
\alpha^{k}
\right].
\end{equation}
The energies $\varepsilon_n^{\rm PDC}$ are shown in
Fig.~\ref{fig:PDCenergies}. It is instructive to compare this figure
with Fig.~\ref{fig:energies} for the energies (\ref{sequence}). The
overall band structure is similar, but the energy distribution is much
more regular for the PDC. For example, it is seen from the panels for
2000 sites that the minibands for the PDC sequence have approximately
equal numbers of states, in contrast to
Fig.~\ref{fig:energies}. However, as we show later, the symmetry of
the PDC is actually bad from the viewpoint of many-particle
localization.

Spatial separation of sites with close energies in the PDC leads to
effective localization of single-particle stationary states. As in the
case of sequence (\ref{sequence}), the values of $\langle
I_1\rangle-1$ are $\sim 3\times 10^{-3}$ for $h/J=20$ near the minimum
of $\langle I_1\rangle$ over $\alpha$, in agreement with the estimate
(\ref{nearest}). The minimum of $\langle I_1\rangle$ is located at
$\alpha\approx 0.1$.

The situation is different for many-particle localization. Here the
high symmetry of the PDC leads to multiple many-particle resonances.
For example, two-particle states $(n,n+1)$ and $(n-1,n+2)$, which are
coupled in second order in $J/h$, have equal energies whenever
$n$ is odd.  The states $(n,n+1)$ and $(n-2,n+3)$, that are coupled in
the fourth order in $J/h$, have the same energies when $n=7+4k$ with
integer $k$. We note that there were no exactly degenerate states for
sequence (\ref{sequence}), and the portion of two-particle states
with close energies was much smaller. Therefore it is more complicated
to find a correction to sequence (\ref{PDC}) that would eliminate
many-particle resonances. As a result, unexpectedly, this symmetric
sequence is less convenient from the point of view of strong
localization.

\subsection{Random on-site energies}

The case opposite to the highly symmetric sequence (\ref{PDC}) is when
the on-site energies $\varepsilon_n$ are completely random.  It is
well-known that such randomness leads to single-particle localization
of stationary states in a 1D chain. However, it does not lead to
strong on-site localization of all states, because there is always a
nonzero probability to have neighboring sites with energies that differ by
less than $J$ and therefore are hybridized. As explained in the
Introduction, hybridization is even more likely to happen in the case
of many-particle states, because it is more likely to have neighboring
nearly resonant sites.

\begin{figure}[h]
\includegraphics[width=3.2in]{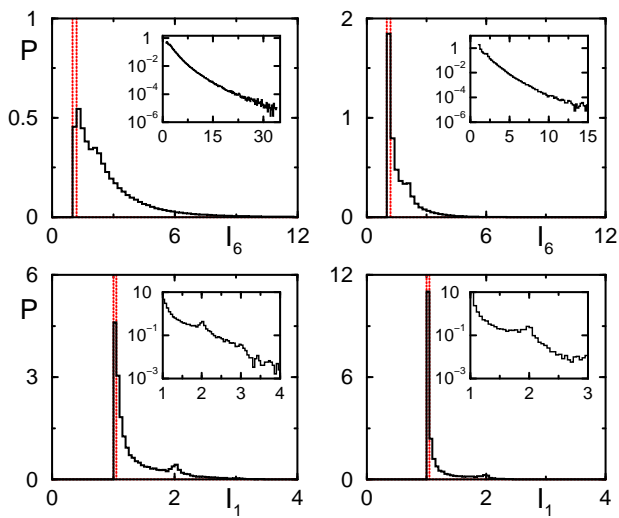}
\caption{(color online) The normalized distributions of the IPR
$P(I)$ for the random energy sequence (\protect\ref{random}). The
upper and lower panels refer to the many- and single-particle
IPR's: 6 excitations with $\Delta =1$ in the chain of length $L=12$
and one excitation in the chain of $L=300$, with 2000 and 80
realizations, respectively. The left and right panels refer to
the overall bandwidth of the on-site energy spectrum $W/J=13$ and
26.  The narrow columns at $I=1$ showed with dotted lines refer
to sequence (\protect\ref{sequence}), with $h/J=10$ and 20
for the left and right panels, respectively. The values of
$\alpha$ in Eq.~(\protect\ref{sequence}) were chosen so as to
have the same bandwidths of on-site energies as the corresponding
random sequences: in the lower panels $\alpha\approx
(W-h)/W\approx 0.231$ (as in an infinite chain), whereas in the
upper panels $\alpha=0.274$ to allow for a comparatively small
number of sites. The insets show $\log P$ vs. $I$. }
\label{fig:rand}
\end{figure}

A simple random sequence of on-site energies has the form
\begin{equation}
\label{random}
\varepsilon_n^{\rm r}=Wr_n^{\prime},
\end{equation}
where $r_n^{\prime}$ with different $n$ are independent random numbers
uniformly distributed in the interval $(0,1)$, and $W$ is the
bandwidth. The results on the inverse participation ratio for sequence
(\ref{random}) are shown in Fig.~\ref{fig:rand}.

Localization of stationary states for the random sequence
(\ref{random}) can be characterized by the probability distribution of
the IPR $P(I)$. This distribution was obtained by numerically
diagonalizing the Hamiltonian (\ref{hamiltonian_fermions}) for
different realizations of the on-site energies (\ref{random}).

As seen from Fig.~\ref{fig:rand}, both single- and many-particle IPR
distributions display peaks near $I=1$ for random on-site
energies. This indicates that, for the wide energy bandwidths used in
the calculations, most of the stationary states of the system are
strongly localized. However, the distributions are broad and slowly
decay on the tails. This means that many on-site states are strongly
hybridized, that is the stationary wave functions spread over several
sites. This is a consequence of multiple resonances.  The insets in
Fig.~\ref{fig:rand} show that, at least for not too large $I$, the
tails of $P(I)$ are non-exponential.

The typical distribution width decreases with the increasing bandwidth
$W$ of on-site energies. As expected, the distributions of
many-particle IPR's (the upper panels in Fig.~\ref{fig:rand}) are much
broader and their peaks near $I=1$ are much smaller than for the
single-particle IPR's (the lower panels).

The IPR distributions for the random sequence (\ref{random}) differ
dramatically from the distributions for the regular sequence
(\ref{sequence}). The latter are narrow and concentrate in a small
region of $I$ close to $I=1$, for the chosen bandwidths, both for the
single- and many-particle states, cf.
Figs.~\ref{fig:one_excitation}, \ref{fig:many_loc}. This is another
indication of strong single- and many-particle localization for sequence
(\ref{sequence}).

\section{Conclusions}

In this paper we have explored two aspects of the problem of strong
many-particle localization. One is localization of stationary
states. For one-particle states, it has been studied analytically. We
found that the wave functions decay quasi-exponentially and obtained
the bounds on the decay length. The numerical results on strong
localization are in agreement with the theory. 

For many-particle stationary states, the localization has been
analyzed numerically. Such analysis is unavoidably limited to small
chains. For a 12-site chain we have found that, for the on-site energy
sequence (\ref{correction_6}) with the ratio of the single-site energy
bandwidth to the hopping integral $\approx 28$
[$h/J=20,\alpha=0.25,\alpha^{\prime} = 0.22$ in Eq.~(\ref{correction_6})], the
inverse participation ratio differs from its value for the case of
fully localized states by $< 2\%$. This deviation is due to a
small nonresonant admixture of the wave functions of neighboring
sites.

A different approach is based on studying the lifetime of on-site
states. It is sufficient to have a localization lifetime $t_{\rm loc}$
that exceeds the coherence time of the excitations. We have shown that
such $t_{\rm loc}$ can be achieved in a chain of an arbitrary length
and with an arbitrary number of excitations. For the explicit
construction of on-site energies (\ref{correction_6}), resonant
transitions that lead to delocalization require at least 4 virtual
nonresonant steps, and for comparatively weak particle-particle
coupling even 6 virtual transitions. Even in the 4-step case this
leads to the ratio of the delocalization rate to the inter-site
hopping integral $\sim 10^{-5}$ already for the decay length of the
amplitude of virtual transitions $\sim 0.35$, or for $K\sim
0.06$.

The results on localization can be tested with artificial
condensed-matter structures, as it was done in the studies of the
effects of quasiperiodicity \cite{Merlin85} (see also
Refs.~\onlinecite{Sokoloff85,Albuquerque03}). A discussion of
experimental implementations is beyond the scope of this paper. Good
examples are proposed models of a quantum computer; we note, however,
that studying localization does not require operations on qubits, and
therefore does not require a fully operational quantum computer.

In terms of quantum computing, an advantageous feature of sequence
(\ref{sequence}) and its modification (\ref{correction_6}) is that one
radiation frequency can be used to resonantly excite different
qubits. This can be achieved by selectively tuning them to this
frequency without bringing neighboring qubits in resonance with each
other, or by sweeping the frequency of the targeted qubit through the
radiation frequency and having a Landau-Zener-type interstate
transition. A two-qubit gate can be conveniently done by selectively
tuning neighboring qubits in resonance with each other and having a
Landau-Zener type excitation swap \cite{DP01}.  

In many cases localization is also a prerequisite for a projective
measurement. This happens when the measured quantity is the
probability for each qubit to be in the excited state. Often a
measurement is much slower than the time $J^{-1}$ of resonant hopping
to a nearest site; then $t_{\rm loc}$ should exceed the measurement
time. In our approach, localization does not require refocusing
\cite{liquid_NMR}, which is not always easy to implement and which is
sometimes incompatible with slow measurement.

Our energy sequence does not lead to delocalization of stationary
states due to indirect resonant $n\leftrightarrow n+2$ transitions. Such
transitions undermine the approach to quantum computing with
``always-on qubit interaction'' proposed in
Ref.~\onlinecite{Benjamin03}, where next nearest neighbors are tuned
in resonance.  In terms of the localization lifetime, the width of the
energy distribution (\ref{correction_6}) is orders of magnitude smaller
than the width that would give the same $t_{\rm loc}$ in the approach
\cite{Benjamin03}. There the localization lifetime would be $\sim
h/J^2$, where $h$ is the energy difference between nearest
neighbors. This difference must exceed $J$ by a factor $\sim 10^5$ in
order to have the same localization lifetime as in our
approach. Compared to Ref.~\cite{Zhou02}, a potential advantage of our
approach is that the interaction does not have to be ever turned off,
and no multi-qubit encoding is necessary for operating a QC.

The presented scheme can be extended to systems with long-range
coupling. This is another distinction from the approaches
\cite{Zhou02,Benjamin03} that substantially rely on nearest neighbor
coupling. For several proposed QC's the interqubit coupling is dipolar
for a few near neighbors and becomes quadrupolar or falls down even
faster for remote neighbors \cite{Makhlin01,mark,Nakamura03}.
Long-range interaction makes transitions over several sites more
probable.  We leave detailed analysis for a separate paper. Here we
note that, for sequence (\ref{correction_3}) and for single-particle
transitions, not only hopping over 2 or 4 sites is not resonant and
does not lead to delocalization, but even hopping over 6, 8, or 10
sites is nonresonant as well. For all these transitions, the energy
difference is at least $\sim \alpha^2 h$ or $\sim \beta h/4$ [for
sites separated by an odd number of positions, the energy difference
is always large, $\sim h$].

Our results provide proof of principle of strong on-site localization
of all states, independent of the system size and the number of
particles. We have not addressed the question of optimization of the energy
sequence, so that maximal localization lifetime could be obtained for
a minimal bandwidth of on-site energies. For a finite-length chain the
optimization problem can be approached using Eq.~(\ref{sequence}) as
an initial approximation and adjusting energies of several specific
sites.

In conclusion, we have proposed a sequence of on-site energies
(\ref{sequence}) and its modifications (\ref{correction_6}),
(\ref{correction_3}) that result in on-site localization of all
single- and many-particle states of interacting spins or fermions in
an infinite chain. The sequence (\ref{sequence}) has low symmetry,
which allows eliminating resonances between the states to a high order
in the hopping integral. In turn, this leads to a long lifetime of
localized many-particle states. When second-order resonances are
eliminated, the lifetime exceeds the reciprocal hopping integral by 5
orders of magnitude provided the bandwidth of on-site energies is
larger than the inter-site hopping integral by a factor of $\sim
40$. We show that it can be further significantly increased by
eliminating fourth-order resonances. The proposed energy sequence is
stable with respect to errors. The results apply to scalable quantum
computers with perpetually coupled qubits. They show that, by tuning
qubit energies, excitations can be prevented from delocalizing between
gate operations.

\begin{acknowledgments}
We are grateful to D.A. Lidar, L.P. Pryadko, and M.E.~Raikh for
helpful discussions. This work was partly supported by the Institute for
Quantum Sciences at Michigan State University and by the NSF through
grant No. ITR-0085922.
\end{acknowledgments}

\appendix
\section{Exponential decay of the transition amplitude}

In this Appendix we give a rigorous proof of the quasi-exponential
decay of the amplitude $K_n(m)$ (\ref{amplitude}) of the transition
from site $n$ to site $n+m$ and establish bounds on the decay
length. We show that, for sequence (\ref{sequence}), in the limit
of small $\alpha$ and for $m\to \infty$
\begin{equation}
\label{bounds_rigor}
\alpha^{-\nu_{_L}m}\leq K_n(m)(2h/J)^m \leq \alpha^{-\nu_{_U} m}.
\end{equation}
We find that $\nu_{_L}\geq 0.89$ and $\nu_{_U}\leq 1.19$.

In order to simplify notations we introduce dimensionless
energies $\varepsilon_n(\alpha)=2\varepsilon_n/h$. From
Eq.~(\ref{sequence})
\begin{equation}
\label{redefined}
\varepsilon_n(\alpha)=(-1)^n-\sum_{k=2}^{n+1}
(-1)^{\left\lfloor n/k\right\rfloor} \alpha^{k-1}.
\end{equation}
We also set $J/2h=1$. Then  $K_n(m)=1/|Q_n(m)|$, where
\begin{equation}
\label{Q_n(m)}
Q_n(m)=\prod_{s=1}^m\left[\varepsilon_{n+s}(\alpha)-
\varepsilon_n(\alpha)\right].
\end{equation}
From Eq.~(\ref{redefined}), $Q_n(m)$ is a polynomial in $\alpha$.

For a polynomial $P(\alpha)$ we define by $\low P(\alpha)$ the
multiplicity of the root $\alpha=0$, i.e., the lowest power of
$\alpha$ in the polynomial. The exponent $\nu$ that characterizes
the decay of $K_n(m)$ (\ref{result}) is given by $\nu = \low Q_n(m)/m$ for
$m\to \infty$.

The data of the numerical experiments presented in Fig.~\ref{fig:nu}
show that $0.894 < \low Q_n(m)/m < 1.12$ independent of $n$ for large
$m$.

To obtain an analytical estimate we rewrite  Eq.~(\ref{Q_n(m)}) as
\begin{equation}
\label{lowdeg_sum}
\low Q_n(m)=\sum^m_{s=1} \low [\varepsilon_{n+s}(\alpha) -
\varepsilon_n(\alpha)].
\end{equation}
Each term $\low [\varepsilon_{n+s}(\alpha) -
\varepsilon_n(\alpha)]$ is an integer between $0$ and $n+1$.

In order to find bounds for $\low Q_n(m)$ we will estimate how
many terms $\low [\varepsilon_{n+s}(\alpha) -
\varepsilon_n(\alpha)]$ exceed a given $i$, $0\leq i\leq n+1$.
For each $i$ we have a subset $S_{nm}(i)$ of the values $s$ that
satisfy this condition,
\begin{equation}
\label{S(i)}
S_{nm}(i)= \{s|1\leq s \leq m,\ \low [\varepsilon_{n+s}(\alpha)
- \varepsilon_n(\alpha)] > i\}.
\end{equation}
The number of elements in $S_{nm}(i)$ is denoted by $h_{n m}(i)$. This
is the number of polynomials $\varepsilon_{n+s}(\alpha) -
\varepsilon_n(\alpha)$ whose expansion in $\alpha$ starts with
$\alpha^k$ with $k>i$. In what follows for brevity we drop the
subscripts $n,m$ and use $S(i)$ and $h(i)$ for $S_{nm}(i)$ and $h_{nm}(i)$.

It follows from the definition that
$$h(0)\ge h(1) \ge h(2)\ge\dots\ge h(n).$$
By construction
\begin{eqnarray}
\label{Q_in_h}
\low
Q_n(m)&=&\sum_{i=1}^n i[h(i-1)-h(i)]\nonumber \\
&&+ (n+1)
h(n) =\sum\nolimits_{i=0}^n h(i)
\end{eqnarray}

From Eq.~(\ref{Q_in_h}) we see that the upper and lower bounds on
$\low Q_n(m)$ are given by the  sums of the upper and lower
bounds of $h(i)$.

In what follows we will use the standard notations: $\liminf$
($\limsup$) means the lower (upper) limit of a sequence,
$\lcm\{i_1,\dots,i_r\}$ is the least common multiple of integers
$i_1,\dots,i_r$, and $\grcd\{i_1,\dots,i_r\}$ is the greatest common
divisor of $i_1,\dots,i_r$. We will also denote by
$[\varepsilon(\alpha)]_k$ the coefficient at $\alpha^k$ in the
polynomial $\varepsilon(\alpha)$, i.e.,
\begin{equation}
\label{coeff_i}
\varepsilon_n(\alpha)=\sum_{k=0}^n[\varepsilon_n(\alpha)]_k\alpha^k.
\end{equation}

\subsection{Lower bound}

In this section we obtain the lower bound of $\low Q_n(m)$. The main
statement is the following lemma.
\begin{lem}\label{low-bound} The lower bound has the form
\[\liminf_{m\to\infty} \frac{\low Q_n(m)}{m} \ge 0.89.\]
\end{lem}

\noindent{\it Proof.}  Consider first the constant term
$[\varepsilon_n(\alpha)]_0$ in Eq.~(\ref{coeff_i}). By definition,
$[\varepsilon_n(\alpha)]_0 = [\varepsilon_{n+2}(\alpha)]_0$, and
$$\low (\varepsilon_{n+s}(\alpha) - \varepsilon_n(\alpha)) =
  \begin{cases}
    0 & \text{for odd\ } s, \\
    \geq 1 & \text{for even\ } s.
  \end{cases}
$$
Hence, we immediately obtain a simple lower bound $\low Q_n(m)\ge
\left\lfloor m/2\right\rfloor$ for large $m$ (in what follows we
always imply $m\to \infty$).

We compute $h(0),\ h(1)$, etc, using that the coefficients
$[\varepsilon_n(\alpha)]_i$ are periodic in $n$ with period
$2(i+1)$. Indeed, from Eq.~(\ref{redefined}),
\begin{eqnarray*}
[\varepsilon_n(\alpha)]_i&=& (-1)^{\left\lfloor n/(i+1)\right\rfloor}\\
&& =(-1)^{\left\lfloor [n+2(i+1)]/(i+1)\right\rfloor}=
[\varepsilon_{n+2(i+1)}(\alpha)]_i.
\end{eqnarray*}
Therefore, the sets of
coefficients $\{[\varepsilon_n(\alpha)]_0,
[\varepsilon_n(\alpha)]_1,\ldots, [\varepsilon_n(\alpha)]_i\}$ are
also periodic in $n$, but with the period $T_i = 2 \lcm\{2, 3,\ldots,
i+1\}$. This is illustrated by the table

$$
\begin{array}{clrrrrrrrr}
\varepsilon_1(\alpha) &= &-1 & -\alpha &  &  & & & &\\
  \varepsilon_2(\alpha) &= & 1 & +\alpha & -\alpha^2 & & & & &\\
  \varepsilon_3(\alpha) &= &-1 & +\alpha & +\alpha^2 & -\alpha^3 & & & &\\
  \varepsilon_4(\alpha) &= &1 & -\alpha & +\alpha^2 & +\alpha^3 & -\alpha^4 & & &\\
  \varepsilon_5(\alpha) &= &-1 & -\alpha & +\alpha^2 & +\alpha^3 & +\alpha^4 & -\alpha^5 & & \\
  \varepsilon_6(\alpha) &= & 1& +\alpha& -\alpha^2& +\alpha^3& +\alpha^4& +\alpha^5& -\alpha^6 & \\
  \varepsilon_7(\alpha)&= & -1 & +\alpha & -\alpha^2 & +\alpha^3 & +\alpha^4 & +\alpha^5 & +\alpha^6 & -\alpha^7
\end{array}
$$

In order to estimate $h(i)$ we need two technical statements.

\begin{lem}\label{arith-progr1} Let $a_0, k, T$ be any integers such that $2k$ does not divide $T$.
Consider any $\;2k/\grcd\{T,2k\}\;$ consecutive elements of an arithmetic
progression $a_j = a_0 + j\cdot T$, and set $\; b_j=
\left\lfloor a_j/k \right\rfloor\!\! \mod 2$.

Then, at least $\left\lfloor k/\grcd\{T,2k\}\right\rfloor$
integers $\,b_j$ are equal to $0$, and at least the same number of $\,b_j$
are equal to $1$.
\end{lem}

\noindent{\it Proof.} Since $2k$ does not divide $T$, sequence
$a_j\!\mod{2k}$ is cyclic in the interval $[0,2k-1]$. This sequence
contains exactly $2k/\grcd\{T,2k\}$ distinct elements. On average,
half of them (at least $\left\lfloor p/\grcd\{T,2k\}\right\rfloor$)
are less than $p$, and another half (the same number) are larger or
equal than $p$. This means that there are at least $\left\lfloor k
/\grcd\{T,2k\}\right\rfloor$ integers $b_j$ that are equal to $0$
and at least the same number of $b_j$ that are equal to $1$.
Q.E.D.

\hfill

The next statement is a corollary of lemma~\ref{arith-progr1} and we
skip the proof.

\begin{cor}\label{cor-equi}
Let $2k$ does not divide $T$. Consider $p$ consecutive elements of
the arithmetic progression $a_j = a_0 +j\cdot T$ and set
$b_j=\left\lfloor a_j/k\right\rfloor\!\!\mod 2$.

Then at least
$\left\lfloor k/\grcd\{T,2k\}\right\rfloor \left\lfloor p\cdot
\grcd\{T,2k\}/2k\right\rfloor$ integers $b_j$ are equal to
$0$, and at least the same number of $b_j$ are equal to $1$.
\end{cor}

We are now in a position to finish the proof of
Lemma~\ref{low-bound}. We notice first that, for $n=a_j$ in the
expression (\ref{redefined}), the coefficient $b_j$ for given $k$
determines the sign of the term $\alpha^{k-1}$ in
$\varepsilon_n(\alpha)$, that is
$(-1)^{b_j}=[\varepsilon_n(\alpha)]_{k-1}$. The number $h(i)$ gives
the probability that, for all $k\leq i$, the polynomial
$\varepsilon_{n+s}(\alpha)$ has the same $b_j$ as
$\varepsilon_n(\alpha)$.

We will now estimate $h(i)$ with $i=1,\ldots,4$ and start with
$h(1)$. We note that $s\in S(1)$ if and only if $s\in S(0)$ and
$[\varepsilon_{n+s}]_1=[\varepsilon_n]_1$. The second condition means
that $\lfloor s/2\rfloor\! \mod\ 2 =0$. By construction (\ref{S(i)}),
for $m\to \infty$ the set $S(0)$ is formed by all numbers $s$ of the
same parity as $n$. This means that $S(0)$ is an arithmetic
progression with period $T_0=2$.  We take $p$ consecutive elements
$s_1,\dots s_p$ of it and use Corollary~\ref{cor-equi} with $k=2$,
because we are interested in the coefficients
$[\varepsilon_{n+s_i}]_1=(-1)^{\lfloor (n+s_i)/2\rfloor}$ in
(\ref{redefined}). By Corollary~\ref{cor-equi}, since $T_0$ is not
divisible by $2k$, for at least $\lfloor p/2\rfloor$ subscripts $s_i$
the coefficients $[\varepsilon_{n+s_i}(\alpha)]_1=1$, and
$[\varepsilon_{n+s_i}(\alpha)]_1=-1$ for at least the same amount of
subscripts $s_i$, i.e., approximately half of the coefficients
$[\varepsilon_{n+s_i}(\alpha)]_1$ coincide with
$[\varepsilon_n(\alpha)]_1$.  Hence, $h(1)\ge h(0)/2$ as
$m\to\infty$. Substituting $h(0)=m/2$ we obtain $h(1) =m/4$.

Similar arguments can be applied to estimate $h(2)$. This requires
finding a portion of the set $S(1)$ which forms $S(2)$. The set $S(1)$
is a nonempty disjoint union of arithmetic progressions with period
$T_1=4$. We will apply Corollary~\ref{cor-equi} with $k=3$ to each of
these progressions and use $p=h(1)$. This gives
$h(2)\ge\left\lfloor 3/2\right\rfloor\left\lfloor
(1/3)(m/4)\right\rfloor =
\left\lfloor m/12\right\rfloor$, or $h(2)/m \geq 1/12$ for $m
\to \infty$.

In the same way we obtain $h(3)/m\geq 1/24$ and $h(4)/m \geq 1/60$ as
$m\to\infty$. Therefore
\[
\low Q_{n}(m)/m \geq
\frac{1}{2}+\frac{1}{4}+\frac{1}{12}+\frac{1}{24}+\frac{1}{60}\ge
0.89,
\]
which finishes the proof of the lower bound.

\subsection{Upper bound}

We start with the proof of the following rough estimate:

\begin{lem}\label{upest} An upper bound has the form
\[\limsup_{n\to\infty} \frac{\low Q_n(m)}{m}\leq \frac{22}{15} <
1.47. \]
\end{lem}

\noindent{\it Proof.} Taking in the rhs of Eq.~(\ref{Q_in_h}) the sum
from $0$ to $\infty$ we obtain
\begin{equation}\label{lowdeg2}
\low Q_n(m)\leq  \sum_{i=0}^\infty h(i).
\end{equation}

To find an upper bound on $h(i)$ we will use the following consequence
of Lemma~\ref{arith-progr1}:

\begin{cor}
\label{cor-upper}
Let $2k$ does not divide $T$. Consider $p$ consecutive elements of
the arithmetic progression $a_j = a_0 +j\cdot T$ and set
$b_j=\left\lfloor a_j/k\right\rfloor\!\!\mod 2$.

Then at most
\begin{eqnarray*}
&&\left({2k\over \grcd\{T,2k\}}-
\left\lfloor {k\over \grcd\{T,2k\}}\right\rfloor \right)\\
&& \times\left(\left\lfloor {p\cdot
\grcd\{T,2k\}\over 2k}\right\rfloor +1\right)
\end{eqnarray*}
integers $b_j$ are equal to $0$, and at
most the same number of $b_j$ are equal to $1$.
\end{cor}

Using the same arguments as before, by corollary~\ref{cor-upper} we
obtain for $m\to \infty$ the following upper bounds for $h(i)$
%

\begin{eqnarray}
\label{h0-4}
  h(0)&\leq& m/2,\nonumber \\
  h(1)&\leq& m/4,\nonumber\\
  h(2)&\leq& m/6,\nonumber\\
  h(3)&\leq& m/12,\nonumber\\
  h(4)&\leq& m/20.
\end{eqnarray}

Recall that $h(4)\geq h(5)\ge h(6)\geq h(7).$ Similarly, for $ q\geq
2$ we have
$$h(2^q)\geq h(2^{q}+1)\geq\dots
h(2^{q+1}-1).$$
Therefore we can replace  the terms
\newline $h(2^q+1),\dots,h(2^{q+1}-1)$ in the rhs of
Eq.~(\ref{lowdeg2}) by $h(2^q)$, which leads to the following upper
bound for $\low Q_n(m)$,
\begin{equation}\label{uppbound1}
  \low Q_n(m)\leq
  h(0)+h(1)+h(2)+h(3)+\sum_{q=2}^\infty 2^q
  h(2^q).
\end{equation}
This reduces the calculation to finding upper bounds on $h(2^q)$.

We will now obtain a recurrence relation for $h(2^q)$. First we notice
that, for $K+1$ being a prime number, we have from Corollary~\ref{cor-upper}
\begin{equation}
\label{prime_increment}
h(K+1)\leq {K +2\over 2(K+1)}h(K).
\end{equation}
For all primes $K+1\geq 7$ we have $(K+2)/2(K+1)\leq 4/7$. Therefore
$h(K+1)\leq {4\over 7}h(K)$ for $K\geq 6$. We also note that, for all
positive integral $q$,
\[h\left(2^{q}\right) \leq {1\over 2} h\left(2^{q}-1\right).\]

Now we recall the distribution law for primes in the intervals.
The following statement is called Bertrand's postulate (or
Tchebychev' theorem) (see Ref.~\onlinecite{N}):
\begin{thm}\label{Tcheb}
There is at least one prime between $M$ and $2M$ for any positive
integer $M$. If $M>3$, there is always at least one prime between
$M$ and $2M-2$.
\end{thm}

In particular, there is at least one prime between $2^q$ and
$2^{q+1}-1$ for any positive integer $q\geq 2$. With this statement,
taking into account the previous estimates, we obtain
\[h(2^{q+2})\leq
 \left( {1\over 2}\cdot{4\over 7} \right)^q {m\over 20} \]
for all positive integral $q$, or
\begin{equation}
\label{hmn2k}
  h(2^{q+2})\leq \left( {2\over 7} \right)^q {m\over 20}.
\end{equation}

Substituting inequalities~(\ref{h0-4}), (\ref{hmn2k}) into
~(\ref{uppbound1}), we obtain
\begin{widetext}
\begin{eqnarray*}
\low Q_n(m)&\leq & {m\over 2} + {m\over 4} + {m\over 6} + {m\over 12}
+{m\over 20}\left(2^2 +2^3\cdot {2\over 7} +2^4\left( {2\over 7} \right)^2+
\ldots +2^k\left(
{2\over 7} \right)^{k-2} +\ldots\right)\\ &=& m\left(1+
{2^2\over 20}\sum_{j=0}^\infty\left( {4\over 7} \right)^j\right) \qquad \text{ for
} \quad m\to \infty.
\end{eqnarray*}

\end{widetext}

This gives
\begin{equation}\label{upper2}
  \low Q_n(m)\leq \frac{22}{15}m < 1.47m
\end{equation}
The last inequality is an explicit asymptotic upper bound.
Q.E.D.\hfill

We now provide a sharper upper bound. We will use the same method, but
instead of Tchebyshev' theorem we will apply Erd\"os theorem.

\begin{lem}\label{lem-sharper}(Sharper bound)
$$\limsup_{m\to\infty}\frac{\low Q_n(m)}{m}\leq 1.19$$
\end{lem}

\noindent{\it Proof.} Following the same pattern as in
Lemma~\ref{upest} above we extend the explicit list of
inequalities~(\ref{h0-4}).

As $m\to\infty$,
\begin{eqnarray}
  h(5)/m &\leq& 1/20,\nonumber \\
  h(6)/m &\leq& 1/35,\nonumber \\
  h(7)/m&\leq& 1/70,\nonumber \\
  h(8)/m&\leq& 1/126,\nonumber \\
  h(9)/m&\leq& 1/126,\nonumber\\
  h(10)/m&\leq& 1/231, \nonumber \\
  h(11)/m&\leq& 1/231, \nonumber \\
  h(12)/m&\leq& 1/429, \nonumber \\
  h(13)/m&\leq& 1/429, \nonumber \\
  h(14)/m&\leq& 1/429, \nonumber \\
  h(15)/m&\leq& 1/858, \nonumber \\
  h(16)/m&\leq& 3/4862.
\label{h5-16}
\end{eqnarray}

To obtain a sharper upper bound we recall the following result by
Erd\"os \cite{E}.

\begin{thm}\label{erdos}(Erd\"os) There exist at least one prime of the form $4k +1$
and at least one prime of the form $4k + 3$ between $M$ and $2M$
for all $m > 6$.
\end{thm}

For all primes that exceed 16 we have in Eq.~(\ref{prime_increment})
$(K+2)/2(K+1) < 7/13$. Therefore, by reproducing the arguments that
led to the inequality (\ref{hmn2k}), but using now the relation
(\ref{prime_increment}) twice based on the theorem~\ref{erdos}, we
obtain
\begin{equation}\label{hmn2k+4}
  h(2^{q+4})\leq
 \left[ \frac{1}{2}\cdot \left(\frac{7}{13} \right)^2 \right]^q h(16).
\end{equation}

Substituting inequalities~(\ref{h0-4}), (\ref{h5-16}), (\ref{hmn2k+4})
into ~(\ref{uppbound1}) (where now the terms up to $h(15)$ are taken
into account explicitly, and the sum runs from $q=4$) we obtain

\begin{eqnarray*}
&&\limsup_{m\to\infty}\ \frac{\low Q_{n,m}}{m} \leq  \frac{1}{2} +
\frac{1}{4} + \frac{1}{6} + \frac{1}{12} + \frac{2}{20}\\
&&+ \frac{1}{35} +
\frac{1}{70} + \frac{2}{126} + \frac{2}{231} + \frac{3}{429} +
\frac{1}{858}\\
&&+
\frac{3\cdot 16}{4862}\left(1+\left( \frac{
7}{13} \right)^2+ \left(  \frac{ 7}{13}
\right)^4 + \ldots
\right).
\end{eqnarray*}

Evaluating the rhs, we obtain
\begin{equation}\label{upper3}
  \limsup_{m\to\infty}\frac{\low Q_n(m)}{m} < 1.19.
\end{equation}
Q.E.D.\hfill

\section{The interrelation between the energy spectrum
parameters $\alpha$ and  $J/h$ for fixed IPR}

In this Appendix we outline another way of looking at the effect of
the band structure of sequence (\ref{sequence}) on
localization. It applies to single-particle stationary states and is
based on varying $h/J$ and finding such energy spectrum parameter
$\alpha$ that would keep the IPR constant, i.e.,
\begin{equation}
\label{I_1const}
\langle I_1\rangle \equiv \langle I_1(\alpha,J/h)\rangle={\rm const}.
\end{equation}

The average IPR is large, $\approx L/3$, when the spread of the
on-site energies $\alpha h$ is small compared to the
hopping-induced bandwidth $J^2/2h$ of the bands at $\pm h/2$.
When $\alpha h$ becomes comparable to $J^2/h$, a part of the
states become localized with localization length smaller than the
chain size, but still there remain states of size $\sim L$. For
such states $I_{1\lambda}\propto L$. Their portion depends on
$\alpha h/(J^2/h)$. Therefore one may expect that, for large
$\langle I_1\rangle$ and for a given chain length, $\alpha$
should vary with $J/h$ as $(J/h)^2$.

Another scaling region of $\alpha(J/h)$ as given by
Eq.~(\ref{I_1const}) may be expected to emerge for $\alpha$ close to
the threshold value, $\alpha_{\rm th}\lesssim \alpha\ll 1$, but far
away from the strong-localization range of $\alpha$, where $\langle
I_1\rangle -1\sim J^2/h^2$ [cf. Eq.~\ref{nearest})].

For $\alpha$ close to $\alpha_{\rm th}$, the wave functions have
comparatively small-amplitude tails that spread over a long distance
and are nearly exponential at large distances, as given by
Eq.~(\ref{result}). If the decay were purely exponential, i.e., the
tail of the wave function centered on site $n$ were of the form
$\psi_{n+m}=K^{|m|}\psi_n$, we would have $\langle I_1\rangle-1\approx
4|K|^2$ for $K\approx J/2\alpha h\ll 1$. From (\ref{I_1const}), this
condition gives scaling $\alpha \propto J/h$. The nonexponential decay
of the wave functions at small to moderate distances (numerically, for
$|m|\sim 4-8$) leads to deviations from this simple scaling.

\begin{figure}[h]
\includegraphics[width=3.2in]{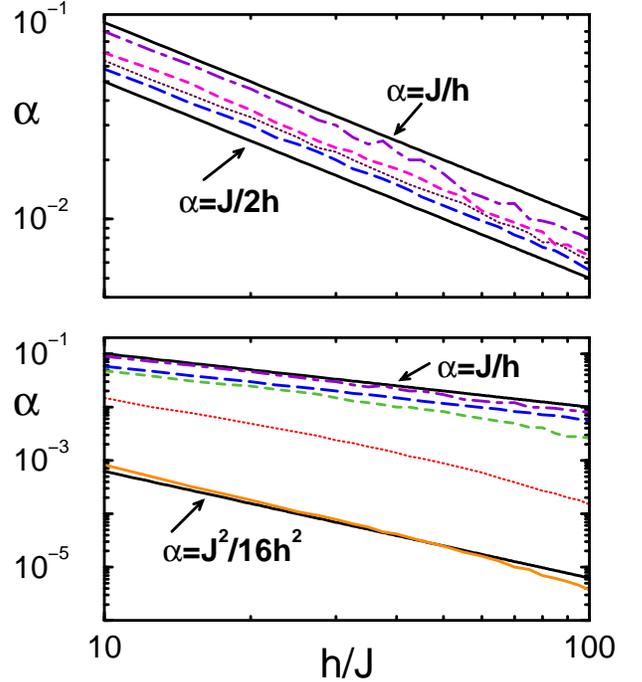}
\caption{(color online). The dependence of $\alpha $ on $h/J$ as given
by the condition $\langle I_1\rangle =$ const for different $\langle
I_1 \rangle$ in the chain with $L=300$. The lines in the lower panel
listed from down upward (thin solid, dotted, dashed, long-dashed, and
dot-dashed) correspond to $\langle I_1\rangle = 95, 50, 10, 2,$ and
1.2, respectively.  The lines in the upper panel listed from down
upwards (long-dashed, dotted, dashed, and dot-dashed) correspond to
comparatively small $\langle I_1\rangle = 2, 1.6, 1.4,$ and 1.2,
respectively. The bold lines $\alpha = J^2/16h^2$ and $\alpha = J/h$
display the asymptotic behavior of $\alpha$ for large and small
$\langle I_1\rangle$; the line $\alpha = J/2h$ corresponds to
$\alpha=\alpha_{\rm th}$.}
\label{fig:alphaXh}
\end{figure}

Numerical results on the dependence of $\alpha$ on $J/h$ as given by
Eq.~(\ref{I_1const}) are shown in Fig.~\ref{fig:alphaXh}. The data for
$\langle I_1\rangle \sim L/3$ show the expected scaling $\alpha\propto
(J/h)^2$. On the other hand, in the range $\langle I_1\rangle-1\approx
$~0.1---1 the value of $\alpha$ scales as $J/h$.  This scaling applies
only for $\alpha > \alpha_{\rm th}$, i.e., for $\alpha h/J > 1/2$. The
value of $\alpha h/J$ as given by Eq.~(\ref{I_1const}) increases with
decreasing $\langle I_1\rangle$.

We note that, for large $h/J\sim 100$ and small $\langle
I_1\rangle-1$, the IPR $\langle I_1\rangle$ as a function of $\alpha$
displays small oscillations. This leads to multivaluedness of the
roots $\alpha$ of the equation $\langle I_1\rangle =$~const. The roots
are numerically very close to each other. We showed the
multivaluedness schematically by plotting $\alpha$ vs. $h/J$ in
Fig.~\ref{fig:alphaXh} with jagged lines.

In the intermediate range of $\langle I_1\rangle$, the function
$\alpha(J/h)$ crosses over from one type of the limiting behavior to
the other. The numerical data does not seem to suggest that $\langle
I_1\rangle$ has a universal scaling form of a function of
$\alpha^{\nu}/(J/h)$ for all $\alpha,\, J/h\ll 1$.

\section{Narrow resonances of the many-particle IPR as function of the
parameter $\alpha$}

In this Appendix we discuss the positions and widths of the narrow
peaks of the IPR seen in Fig.~\ref{fig:many_loc}.

As we increase $\alpha$ starting from $\alpha=0$, pronounced peaks of
$\langle I_6\rangle$ appear for the difference in the combination
two-particle on-site energies $\delta\varepsilon=
|\varepsilon_{k_1}+\varepsilon_{k_2}-\varepsilon_{k_3}-\varepsilon_{k_4}$
(\ref{pair_diff}) equal to
\[ \delta\varepsilon \approx s\alpha h
\approx J\Delta\]
with $s=1,2$. They are due to resonant hybridization of pairs on
neighboring sites $(n,n+1)$ with dissociated pairs located on sites
$(n,n+3)$ for $s=1$, and $(n-1,n+2)$ with even $n$ for $s=2$, for
example. Such hybridization corresponds to two single-particle steps
by one site, i.e. $\varkappa = 2$.

A specific example for
the studied chain with $\Delta = 1$, $h/J=20$, and $\alpha=0.05$ is
the resonance between the on-site states $|\Phi(3,4,6,7,8,9)\rangle$,
$|\Phi(4,5,6,7,8,9)\rangle$, $|\Phi(3,4,6,7,8,11)\rangle$ and
$|\Phi(4,5,6,7,8,11)\rangle$ (we remind that the arguments of $\Phi$
indicate the positions of the excitations; we have six excitations,
and the available sites are $1,2,\ldots,12$). All these states can be
obtained from each other by moving one excitation by two
positions. For example, in the first pair the excitation goes from
site 3, where it has one nearest neighbors, to site 5, where it has
two neighbors.

The width of the above peaks $\delta\alpha$ can be estimated from the
condition that the frequency detuning $|s\alpha h\pm J\Delta|$ is of
order of the effective hopping integral $J\Delta
V_{k_1k_2k_3k_4}$. For $s=1$ [an $(n,n+1)\leftrightarrow (n,n+3)$-type transition]
the hopping integral is $\sim J^3\Delta/\alpha h^2$ from
Eqs.~(\ref{H_int}), (\ref{V4}). This gives the width
\[\delta\alpha\sim (J/h)^2.\]
The positions of the peaks $\alpha \approx J\Delta/sh$ and
their widths are in agreement with the data in both upper and lower
main panels of Fig.~\ref{fig:many_loc}.

For larger $\alpha$, narrow resonances with respect to $\alpha$ occur when
\[s\alpha^mh\approx MJ\Delta\]
with integer $s,m,M$, and $m\geq 2, M=1,2$. They may happen, for
example, between pairs $(n,n+1)$ and $(n-1,n+2)$ with odd $n$ such
that $n\neq 3k-1$, in which case $m=2$ and $\varkappa =2$. A specific
example for our chain is the resonance between the on-site states
$|\Phi(1,3,4,6,9,11)\rangle$ and $|\Phi(1,2,5,6,9,11)\rangle$ for
$\Delta =1, h/J=20$, and $\alpha=0.246$. Here the excitations on sites
$(3,4)$ move to sites $(2,5)$, and $\alpha^2h\sim J\Delta$ (in fact,
higher-order terms in $\alpha$ are essential for fine-tuning the
states into resonance).  In other cases resonances with $m \geq 2$
require more intermediate virtual steps, with $\varkappa \geq 4$.

The $m\geq 2$-resonances are extremely narrow for $\alpha_{\rm th}\ll
\alpha\ll 1$. For example, for $m=2$ their widths are
\begin{eqnarray*}
\begin{array}{lll}
\delta\alpha \lesssim
(J/h)^{5/2}\Delta^{1/2} & {\rm for} & \varkappa=2,\\
\delta\alpha \lesssim J^3/h^3\Delta  &
{\rm for} & \varkappa=4.
\end{array}
\end{eqnarray*}
In these estimates we used that, from Eqs.~(\ref{sequence}),
(\ref{V4}) $|V_{n-1\,n+2\,n\,n+1}|\lesssim J^2/h^2$ for the
$\varkappa=2$-transition $(n,n+1) \leftrightarrow (n-1,n+2)$. For the $m=2$ and
$\varkappa=4$-transitions, on the other hand,
$|V_{k_1k_2k_3k_4}|\lesssim J^4/\alpha^3 h^4$ [for example, this
estimate applies to a transition $(n,n+1)\leftrightarrow (n,n+5)$]. We note that,
from the condition $\alpha_{\rm th}\ll \alpha$ and the resonance
condition $s\alpha^2h=MJ\Delta$, it follows that $\Delta\gg J/h$,
which guarantees the smallness of the peak widths.

Each high-order resonance gives rise to a narrow band of resonant
$\alpha$ values. All of them refer to a resonant transition
between the same sites. However, the energy difference of these
sites is slightly different depending on the occupation of remote
sites, for example, next nearest neighbors. In this latter case,
from Eqs.~(\ref{H_int}), (\ref{V4}), the corresponding shift of
$\alpha$ is $\propto (J/h)^{5/2}\Delta^{1/2}$.

A specific example for the studied chain is provided by the resonances
between two pairs of on-site states, $|\Phi(2,3,6,7,8,12)\rangle$ and
$|\Phi(2,3,6,8,11,12)\rangle$, on the one hand, and
$|\Phi(5,6,7,8,9,12)\rangle$ and $|\Phi(5,6,8,9,11,12)\rangle$, on the
other hand. In both cases the resonant transition is fermion hopping
from site 7 to site 11. Both resonances occur for $\Delta = 1,
h/J=20$, but the first corresponds to $\alpha \approx 0.2778$, whereas
the second corresponds to $\alpha \approx 0.2782$.  The difference in
$\alpha$ comes primarily from the different occupation of the next
nearest neighbors of sites 7 and 11.

The most pronounced peaks in Fig.~\ref{fig:many_loc} correspond to
comparatively small $\varkappa\leq 4$. However, there are resonances
for higher $\varkappa$ as well. The resonating energies have to be
extremely close to each other for such states to be hybridized. The
corresponding peaks are very narrow, and very high precision
is needed to find them numerically (sometimes the hybridization
appears to be an artifact of not sufficiently precise calculations).

As mentioned above, in the region $0.2\lesssim \alpha\lesssim 0.4$ the
positions of the IPR peaks are determined not only by the
leading-order terms in $\alpha$, but also by higher-order
terms. Therefore there are several resonant bands for each $s,m,M$ as
given by the condition $s\alpha^mh \approx MJ\Delta$. This explains
why there are several bands in Fig.~\ref{fig:many_loc}. These bands
are well separated for sufficiently large $h/J$ and not too large
$\Delta$. On the other hand, for $h/J=10$ and $\Delta =3$ the bands of
resonances are broadened and overlap with each other. When $\alpha$ is
not very small there emerge also narrow resonances where
$\delta\varepsilon \ll J$ and $\delta\varepsilon/h \to 0$. They are
responsible for some of the peaks in the insets of
Fig.~\ref{fig:many_loc}.

\end{document}